\documentclass[12pt]{article}
\usepackage{amsmath,amssymb,amsfonts,epsfig}
\usepackage[tight]{subfigure}

\usepackage{bm}

\makeatletter\@addtoreset{equation}{section}\makeatother

\newcommand{\bep}{\left(\begin{array}{ccc}}
\newcommand{\bepr}{\left(\begin{array}{rrr}}
\newcommand{\eep}{\end{array}\right)}
\newcommand{\beB}{\left\{\begin{array}{ccc}}
\newcommand{\beBr}{\left\{\begin{array}{rrr}}
\newcommand{\eeB}{\end{array}\right\}}
\newcommand{\nn}{\nonumber}
\newcommand{\pp}{\prime\prime}

\newcommand{\tr}{\text{tr}}
\newcommand{\trn}[2]{\text{tr}_{#1}\left( #2 \right)}
\newcommand{\bra}[1]{\left\langle #1 \right|}
\newcommand{\ket}[1]{\left| #1 \right\rangle}
\newcommand{\vev}[1]{\left\langle #1 \right\rangle}

\newcommand{\ylm}{Y_{lm}}
\newcommand{\tlm}{T_{lm}}
\newcommand{\tn}[1]{T_{l_{#1}m_{#1}}}
\newcommand{\tnd}[1]{T_{l'_{#1}\,m'_{#1}}}

\newcommand{\plm}{\phi_{lm}}
\newcommand{\pn}[1]{\phi_{l_{#1}m_{#1}}}
\newcommand{\pin}[1]{\phi^{\text{in}}_{l_{#1}m_{#1}}}
\newcommand{\pind}[1]{\phi^{\text{in}}_{l'_{#1}\,m'_{#1}}}

\newcommand{\phin}{{\phi^{\text{in}}}}
\newcommand{\phinA}{{\phi^{\text{in}}}^A}

\newcommand{\ton}[1]{T_{2L\,m_{#1}}}
\newcommand{\tond}[1]{T_{2L\,m'_{#1}}}
\newcommand{\tom}[1]{T_{2L\,-m_{#1}}}

\newcommand{\pon}[1]{\phi^{\text{out}}_{2L\,m_{#1}}}
\newcommand{\pond}[1]{\phi^{\text{out}}_{2L\,m'_{#1}}}

\newcommand{\phout}{{\phi^{\text{out}}}}

\newcommand{\wphi}{\widetilde{\phi}}
\newcommand{\wL}{\widetilde{L}}
\newcommand{\wT}{\widetilde{T}}

\newcommand{\Lap}[1]{\left[L_i,\left[L_i,#1\right]\right]}

\newcommand{\Lo}{\Lambda_{\text{out}}}
\newcommand{\Li}{\Lambda_{\text{in}}}

\newcommand{\skinin}{S_N^{(\text{kin}.)\,\text{in}}}
\newcommand{\skinout}{S_N^{(\text{kin}.)\,\text{out}}}
\newcommand{\spot}{S_N^{(\text{pot}.)}}

\allowdisplaybreaks

\input epsf

\begin{document}

\thispagestyle{empty}
\vspace*{-2em}
\begin{flushright}
RUP-12-1
\end{flushright}
\vspace{0.3cm}
\begin{center}
\Large {\bf
Renormalization group approach 
to matrix models via noncommutative space}

\vspace{0.7cm}

\normalsize
 \vspace{0.4cm}

Shoichi {\sc Kawamoto}$^{a,}$\footnote{e-mail address:\ \ 
{\tt kawamoto@thu.edu.tw, kawamoto@yukawa.kyoto-u.ac.jp}},
Tsunehide {\sc Kuroki}$^{b,}$\footnote{e-mail address:\ \ 
{\tt tkuroki@rikkyo.ac.jp}}
and
Dan {\sc Tomino}$^{a,}$\footnote{e-mail address:\ \ 
{\tt dantomino@thu.edu.tw}}

\vspace{0.7cm}

$^a$ 
{\it
Department of Physics, Tunghai University, \\
Taichung 40704, 
Taiwan, R.O.C.}\\

\vspace{0.4cm}

$^b$
{\it 
Department of Physics, Rikkyo University, \\
Tokyo 171-8501, Japan.}\\

\vspace{1cm}
{\bf Abstract}
\end{center}
We develop a new renormalization group approach to the large-$N$ limit 
of matrix models. 
It has been proposed that a procedure, in which
a matrix model of size $(N-1) \times (N-1)$
is obtained by integrating out one row and column 
of an $N \times N$ matrix model,
can be regarded as a renormalization 
group and that its fixed point reveals critical behavior in the
large-$N$ limit. 
We instead utilize the fuzzy sphere structure based on which 
we construct a new map (renormalization group) from $N\times N$ matrix model 
to that of rank $N-1$. 
Our renormalization group 
has great advantage of being a nice analog of the standard 
renormalization group in field theory. 
It is naturally endowed 
with the concept of high/low energy, and consequently it is in a sense local 
and admits derivative expansions in the space of matrices. 
In construction we also find that our renormalization in general generates 
multi-trace operators, and that nonplanar diagrams yield a nonlocal
operation on a matrix, whose action is to
transport the matrix to the antipode on the sphere. 
Furthermore the noncommutativity of the fuzzy sphere 
is renormalized in our formalism.
We then analyze our renormalization group equation, and Gaussian
and nontrivial fixed points are found.  
  We further clarify how to read off scaling dimensions from our 
renormalization group equation.
Finally the critical exponent of the model of
two-dimensional gravity based on our formalism is examined.

\newpage
\setcounter{page}{1}

\tableofcontents

\setcounter{footnote}{0}

\section{Introduction}
\label{sec:intro}

There have been a lot of studies on
nonperturbative aspects of string theory that have revealed that 
string theory would be nonperturbatively formulated in terms of
a matrix model or a gauge theory in the large-$N$ limit.
In fact, it is well known that one- or two-matrix models 
provide nonperturbative formulations of noncritical strings \cite{Knizhnik:1988ak} 
defined in less than one dimension \cite{Brezin:1990rb}. 
Furthermore, several 
large-$N$ gauge theories or matrix models are proposed as nonperturbative formulations 
of critical superstring theories \cite{Banks:1996vh}. 
However, the latter models are 
incomplete so far because we do not know precisely 
how to take a double scaling limit
in which the gauge coupling is tuned to its critical value in a
correlated way with the large-$N$ limit.
In fact, in order to specify the recipe of the double scaling limit,
we need information on dynamics of theory 
such as a critical exponent. 
In the case of noncritical strings, we can get it via the orthogonal
polynomial method for a matrix model, or conformal field theory
techniques for continuum (Liouville) theory. 
On the other hand, matrix models or gauge theories proposed as the
models of critical string theories have more complicated actions with multiple matrices
and this fact makes these models difficult to solve exactly,
 by traditional methods 
like the orthogonal polynomials, the Schwinger-Dyson equations,
symmetry argument and so on. 
However, we here notice that in order to extract a critical exponent,
we do not always have to solve a model explicitly. 
Indeed, we know that universal quantities like a critical exponent 
can be often derived by the renormalization group (RG) approach
\cite{Wilson:1973jj}, 
where we do not have to integrate all degrees of freedom in a model. 
In the case of matrix models, this procedure has been materialized 
in a concrete way which is recapitulated in the next subsection.

\subsection{A brief review of large-$N$ renormalization group}
\label{sec:review_BZJ}

We review
a good example in the case of Hermitian one-matrix model presented 
by Brezin and Zinn-Justin in \cite{Brezin:1992yc}. 
For the purpose of examining the critical exponent in the large-$N$ limit, 
they first decompose $N\times N$ matrix $\phi_N$ by 
\begin{align}
\phi_N=
\bep
\phi_{N-1} & v \\
v^\dagger & \alpha
\eep,
\end{align} 
where $\phi_{N-1}$ is an $(N-1)\times (N-1)$ matrix,
$v$ is an $(N-1)$-component vector, and $\alpha$ is a real number. 
Then only $v$ and $\alpha$, namely one row and one column, 
are integrated out to yield a matrix model of
$\phi_{N-1}$.  As illustration, starting from the action
\begin{align}
S_N=N\trn{N}{\frac12\phi_N^2+\frac{g}{4}\phi_N^4}, 
\label{onemm}
\end{align}
then at one-loop level and to the leading order in $1/N$ expansion,
we obtain an action of $(N-1)\times (N-1)$ matrix model, 
\begin{align}
S_{N-1}=N\trn{N-1}{\left(\frac12+\frac{g}{N}\right)\phi_{N-1}^2+\frac{g}{4}\phi_{N-1}^4}.
\label{N-1action}
\end{align}
The key observation here is that we can repeat this procedure and as such this is quite 
analogous to the usual RG. This motivates us to make a ``wave function renormalization"
\begin{align}
\phi_{N-1}=\rho\phi'_{N-1}, \quad \rho=1-\frac{2g+1}{2N}+{\cal O}\left(\frac{1}{N^2}\right),
\end{align}
so that the action \eqref{N-1action} will have again the standard kinetic term 
in the $(N-1)\times (N-1)$ matrix model 
\begin{align}
S_{N-1}=(N-1)\trn{N-1}{\frac12{\phi'}^2_{N-1}+\frac{g'}{4}{\phi'}^4_{N-1}},
\end{align}
from which we read off the change of the coupling constant as 
\begin{align}
g'=g-\frac1N(g+4g^2).
\label{gchange}
\end{align}
Now it is crucial that since $N$ plays a role of the cutoff in the standard RG, 
we assume the Callan-Symanzik like RG equation 
\begin{align}
\left[N\frac{\partial }{\partial N}-\beta(g)\frac{\partial }{\partial g}+\gamma(g)\right]
F_N(g)=r(g),
\label{CS}
\end{align}
where $F_N(g)$ is the free energy of the $N\times N$ matrix model: 
\begin{align}  
F_N(g)=-\frac{1}{N^2}\log Z_N(g), \qquad Z_N(g)=\int d\phi_N\,e^{-S_N}.
\end{align}
Namely, \eqref{CS} prescribes the $N$-dependence of $F_N$. 
By plugging the known exact result, we can check that the free energy 
near the critical point $g_*=-1/12$ in fact satisfies \eqref{CS} 
and there $\beta'(g)$ is related to the critical exponent 
(string susceptibility) $\gamma_1$ as
\begin{align}
\gamma_1=\frac{2}{\beta'(g_*)}. 
\label{exponent}
\end{align}
We expect that at a fixed point of the above RG \eqref{gchange}, 
critical behavior would show up and analysis of the RG near the fixed point 
gives us the critical exponent. This is indeed the case. 
{}From \eqref{gchange} we can read off the $\beta$-function, 
namely dependence of the coupling constant on the cutoff $N$  as 
\begin{align}
\beta(g)=-g-4g^2,
\label{BZbeta}
\end{align}
and we find the nontrivial fixed point $g_*=-1/4$ and 
the critical exponent
$\gamma_1=2$ according to \eqref{exponent}. 
They relatively well approximate
the exact values $g_*=-1/12$ and $\gamma_1=5/2$. 
Hereafter we call this approach as large-$N$ renormalization group (RG).\footnote
{More applications of the large-$N$ RG are presented in e.g. \cite{Higuchi:1993he}.}\\

When we attempt to apply the large-$N$ RG to matrix models for critical strings, 
we immediately recognize the following issue: interpretation of matrices 
in these ``new" matrix models is quite different from that in the ``old" matrix models 
describing noncritical strings. 
Namely, in the former we interpret matrices as carrying information 
on the space-time itself; for example, eigenvalues as the space-time coordinates 
(of D-branes) and off-diagonal components as open strings connecting them. 
This is in contrast to the case of noncritical string where a matrix model 
is interpreted as a tool for describing a discretized worldsheet. 
Hence according to the spirit of the RG \cite{Wilson:1973jj}, 
it would be better to integrate out off-diagonal components far from 
diagonal ones, because they correspond to highly massive modes in the space-time, 
not just integrating one row and column as in \cite{Brezin:1992yc}. 
These considerations tempt us to develop a new large-$N$ RG in which 
we assign the concept of high/low energy to each matrix element, and 
we integrate out modes with the highest energy. 
If it is possible, the new large-$N$ RG 
evidently accords with the spirit of the original RG \cite{Wilson:1973jj} 
and is expected to have nice properties like locality. This is the motivation 
of the present work. We show that we can in fact define a new large-$N$ RG 
in a well-defined manner based on this idea. 
The key is utilizing the fuzzy sphere structure \cite{Hoppe} 
by which each matrix element carries angular momentum. 
As a consequence, our large-$N$ RG has a nice correspondence  
with the usual RG in field theory. 
In formulating our large-$N$ RG, we will also clarify in what sense it is ``local" 
and mention its particular features. We then analyze fixed points of our RG 
and propose how to read off a critical exponent. 

This paper is organized as follows: in the next section 
we formulate our large-$N$ RG on the fuzzy sphere and derive the RG equation concretely. 
Fixed points of the RG equation are analyzed in section  \ref{sec:fp}.
It is desirable that these fixed points 
can be related to critical phenomena which have been
observed by numerical simulations \cite{Martin:2004un, Panero:2006bx,
  Das:2007gm} and an analytical study \cite{Steinacker:2005wj} of
fuzzy sphere $\phi^4$ theory via a matrix model.
Our results will be compared to these works. 
Section \ref{sec:discussions} are devoted to conclusions and discussions. 
In appendices we enumerate useful formulas which are necessary in our calculations 
and show some identities satisfied by the fuzzy spherical harmonics.

\section{Formulation of large-$N$ renormalization group on fuzzy sphere}
\label{sec:RG}
In this section we define a filed theory on fuzzy sphere as a matrix model. 
Then we propose a new large-$N$ RG 
based on the analogy of the angular momentum realized 
in the space of matrices by the fuzzy sphere. 

\subsection{Definition of fuzzy sphere}
\label{subsec:fs}

We begin with constructing a map from the space of functions on the sphere 
to that of $N\times N$ matrices following \cite{Hoppe}. 
A function on $S^2$ with radius $\rho$ can be expanded as 
\begin{align}
\phi(\theta,\varphi)=\sum_{l=0}^{\infty}\sum_{m=-l}^l\plm Y_{lm}(\theta,\varphi),
\label{fdecomp}
\end{align} 
where $\theta$, $\varphi$ are the polar coordinates, and $l$, $m$ correspond 
to the angular momentum, the magnetic quantum number (projection of 
the angular momentum), respectively. $\ylm$ is the spherical harmonics 
\begin{align}
\ylm(\theta,\varphi)=\sqrt{(2l+1)\frac{(l-m)!}{(l+m)!}}
P_l^m(\cos\theta)e^{im\varphi},
\label{ylmdef}
\end{align}
which satisfies the orthogonality and completeness condition\footnote
{For later use we change the normalization of $\ylm$ by $\sqrt{4\pi}$ 
from the standard one.}
\begin{align}
&\frac{1}{4\pi}\int d\Omega\, \ylm^*(\theta,\varphi)\ylm(\theta,\varphi)
=\delta_{ll'}\delta_{mm'}, 
\label{yortho} \\
&\frac{1}{4\pi}\sum_{lm}\ylm(\theta,\varphi)\ylm(\theta',\varphi')
=\frac{\delta(\theta-\theta')\delta(\varphi-\varphi')}{\sin\theta}.
\label{ycomp}
\end{align}
$\ylm(\theta,\varphi)$ can be also expanded as  
\begin{align}
\ylm(\theta,\varphi)=\rho^{-l}\sum_{i_1 \cdots i_l}c^{(lm)}_{i_1 \cdots i_l}
x^{i_1}\cdots x^{i_l},
\label{yexp}
\end{align}
where $x^i$ ($i=1\sim 3$) are the standard flat coordinate of $\bm R^3$, 
and $c^{(lm)}_{i_1 \cdots i_l}$ is traceless and totally symmetric 
with respect to $i_1,\cdots,i_l$. $\ylm^*=(-1)^mY_{l-m}$ implies 
that ${c^{(lm)*}_{i_1\cdots i_l}}=(-1)^mc^{(l\,-m)}_{i_1\cdots i_l}$.

In order to define an $N\times N$ matrix corresponding to $\ylm$, 
we first introduce the spin $L=(N-1)/2$ representation of $SU(2)$, 
$L_i$ ($i=1\sim 3$) which satisfies
\begin{align}
[L_i,L_j]=i\epsilon_{ijk}L_k,
\label{generator}
\end{align} 
and define 
\begin{align}
\hat{x}^i=\alpha L_i, \qquad \alpha=\rho\sqrt{\frac{4}{N^2-1}}.
\label{alphadef}
\end{align}
Thus $N\times N$ matrices $\hat x^i$'s satisfy an analogous equation of $S^2$: 
$\sum_{i=1}^3 \hat x^{i^2}=\rho^2$. 
We then define an $N\times N$ matrix $\tlm$ corresponding to $\ylm$ in \eqref{yexp} by 
\begin{align}
\tlm=\rho^{-l}\sum_{i_1 \cdots i_l}c^{(lm)}_{i_1 \cdots i_l}
\hat x^{i_1}\cdots \hat x^{i_l},
\label{texp}
\end{align}
which we call the fuzzy spherical harmonics hereafter and they are considered 
to form a noncommutative algebra of functions on the fuzzy sphere. 
It follows immediately from the property of $c^{(lm)}_{i_1 \cdots i_l}$ that
\begin{align} 
\tlm^{\dagger}=(-1)^mT_{l\,-m}.
\label{Hermiticity}
\end{align}
Since $\tlm$ is in the same representation as that of $\ylm$ 
under the rotation $SU(2)$ of $S^2$, 
\begin{align}
U(R)\tlm U(R)^{-1}=\sum_{m'=-l}^lT_{lm'}R^l_{mm'}(R),
\label{rep}
\end{align}
where $U(R)$ and $R^l(R)$ are the $N$-dimensional and $(2l+1)$-dimensional 
representation of $R\in SU(2)$, respectively. 
The $N$-dimensional representation space of $\tlm$ ($0\leq l \leq 2L=N-1, \; -l\le m\le l$) 
is labeled by an integer $s$ ($-L\leq s\leq L$). From 
the Wigner-Eckart theorem, \eqref{rep} leads to 
\begin{align}
\bra{s}T_{lm}\ket{s'}
=(-1)^{L-s}
\bep
L & l & L \\
-s & m & s'
\eep
R(N,l),
\end{align}
where $R(N,l)$ is independent of $s$, $s'$ and depends on 
the choice of $\ket{s}$. Following the convention in \cite{Iso:2001mg}, 
we take 
\begin{align}
\bra{s}T_{lm}\ket{s'}
=(-1)^{L-s}
\bep
L & l & L \\
-s & m & s'
\eep
\sqrt{(2l+1)N},
\label{Tdef}
\end{align}
so that 
\begin{align}
&\frac1N\trn{N}{\tlm\tnd{}^{\dagger}}=\delta_{ll'}\delta_{mm'}, 
\label{Tortho}\\
&
\frac1N \sum_{lm} 
\bra{s_1}\tlm\ket{s_2}\bra{s_3}\tlm^{\dagger}\ket{s_4}
=\delta_{s_1s_4}\delta_{s_2s_3},
\label{Tcomp}
\end{align}
which can be derived easily from \eqref{Tdef} by using \eqref{p453(3)} and \eqref{p453(4)}, 
where the trace is taken over an $N\times N$ matrix. 
In particular, \eqref{Tcomp} leads to 
\begin{align}
&\frac1N\sum_{lm}\trn{N}{O_1\tlm}\trn{N}{O_2\tlm^{\dagger}}=\trn{N}{O_1O_2}, 
\label{merging}\\
&\frac1N\sum_{lm}\trn{N}{O_1\tlm O_2\tlm^{\dagger}}=\tr_NO_1
\, \tr_NO_2,
\label{splitting}
\end{align}
for arbitrary $N\times N$ matrices $O_1$, $O_2$. 
\eqref{Tortho} and \eqref{Tcomp} show that $\tlm$
form an orthogonal and complete basis of the space of $N\times N$ matrices. Namely, 
any $N\times N$ matrix $\phi$ can be decomposed as 
\begin{align}
\phi=\sum_{l=0}^{2L}\sum_{m=-l}^l\plm\tlm.
\label{phidecomp}
\end{align}
Comparing \eqref{fdecomp} and \eqref{phidecomp}, we find that 
the space of $N\times N$ matrices can be regarded as a regularized space 
of functions on $S^2$ in which the angular momentum $l$ is cut off at $2L=N-1$. 
It cannot be overemphasized that the space of functions on the fuzzy sphere 
has therefore completely finite dimensions, but that nevertheless they form a closed algebra 
and preserve the rotational symmetry $SO(3)$ as in \eqref{rep}.

By taking the large-$N$ limit with $\rho$ fixed, it is clear from \eqref{alphadef} that 
\begin{align}
[\hat x^i, \hat x^j]=i\alpha\epsilon_{ijk}\hat x^k \quad \rightarrow \quad 0,  
\end{align}
namely the standard commutative $S^2$ will be recovered. More precisely, 
as shown in appendix \ref{app:tlm}, 
the structure constant of the fuzzy spherical harmonics 
\begin{align}
\left[\tn{1},\tn{2}\right]=\sum_{l_3m_3}f_{l_1m_1\,l_2m_2\,l_3m_3}\tn{3}^{\dagger},
\end{align}
tends to that of the usual spherical harmonics in the large-$N$ limit 
with $\rho$ fixed.  

{}From the definition of $\tlm$ \eqref{texp}, we find that 
\begin{align}
\Lap{\tlm}=l(l+1)\tlm,
\label{Casimir}
\end{align}
which implies that $\Lap{\cdot}$ corresponds to the Laplacian 
on the unit $S^2$. In fact, the operator on the unit $S^2$ 
${\cal L}_i=-i\epsilon_{ijk}x_j\partial_k$ satisfies 
\begin{align}
&{\cal L}^2=-\Delta_{\Omega}, \nn \\
&\Delta_{\Omega}=
\frac{1}{\sin\theta}\frac{\partial}{\partial\theta}\sin\theta\frac{\partial}{\partial\theta}
+\frac{1}{\sin^2\theta}\frac{\partial^2}{\partial\varphi^2}.
\end{align}
Hence ${\cal L}_i$ essentially corresponds to the adjoint action of $L_i$. 

In summary of this subsection, we have constructed the mapping rules: 
\begin{enumerate}
\item function $\rightarrow$ matrix:
\begin{align}
\phi(\theta,\varphi)=\sum_{l=0}^{2L}\sum_{m=-l}^l\plm Y_{lm}(\theta,\varphi)
\quad \rightarrow \quad
\phi=\sum_{l=0}^{2L}\sum_{m=-l}^l\plm \tlm.
\end{align}
\item integration $\rightarrow$ trace:
\begin{align}
\int\frac{d\Omega}{4\pi}\phi(\theta,\varphi)=\frac1N\tr_N\phi. 
\label{intandtr}
\end{align}
Notice that this holds as equality. 
\item Laplacian $\rightarrow$ double adjoint action:
\begin{align}
-\Delta_{\Omega}\phi(\theta,\varphi) \quad \rightarrow \quad \Lap{\phi}.
\label{Lapandadj}
\end{align}
\end{enumerate}

\subsection{Field theory on fuzzy sphere}
\label{subsec:ftonfs}
In the previous subsection we have seen that functions on the fuzzy sphere 
are defined as $N\times N$ matrices. Hence considering a field theory 
on the fuzzy sphere amounts to constructing a corresponding matrix model. 
We have identified $N^2$ bases 
$\{\tlm\}$ in the space of 
$N\times N$ matrices, and it is remarkable that they enjoy the concept 
of the angular momentum $(l,m)$ based on which we will formulate a new  
RG in the next subsection. 

In the following we restrict our interest to a Hermitian matrix, 
which corresponds to a real function on $S^2$. \eqref{Hermiticity} leads to 
\begin{align}
\phi=\phi^{\dagger} \quad \leftrightarrow \quad \plm^*=(-1)^m\phi_{l\,-m},
\end{align}
for \eqref{phidecomp}. As an illustration we consider\footnote{%
A coupling constant $m^2$ in front of  $\tr_N\, \phi^2$ 
should not be confused with the magnetic quantum number $m$.
Later, the mass $m^2$ is always accompanied by a subscript $N$, and
may not be confusing.}
\begin{align}
S_N=\frac{\rho^2}{N}\trn{N}{-\frac{1}{2\rho^2}[L_i,\phi]^2+\frac{m^2}{2}\phi^2+\frac{g}{4}\phi^4},
\label{action}
\end{align}
which is a natural regularization of the scalar field theory on $S^2$ with radius $\rho$
\begin{align}
S=\int\frac{\rho^2d\Omega}{4\pi}\left(-\frac{1}{2\rho^2}\left({\cal L}_i\phi(\theta,\varphi)\right)^2
+\frac{m^2}{2}\phi(\theta,\varphi)^2+\frac{g}{4}\phi(\theta,\varphi)^4\right).
\end{align}
Note that this action is classically equivalent to \eqref{action} by \eqref{intandtr} and \eqref{Lapandadj}.  
Expanding $\phi$ as in \eqref{phidecomp} and using the trace formula \eqref{4T1} 
given in appendix \ref{app:tlm}, \eqref{action} can be written 
in the momentum space\footnote{Strictly speaking, we are considering 
an analogy of the angular momentum space, which will be called the momentum space 
for simplicity.} as 
\begin{align}
S_N=&\sum_{lm}\frac12(l(l+1)+\rho^2m^2)\plm^*\plm \nn \\
&+\frac{N\rho^2g}{4}\sum_{{l_rm_r}\atop{(r=1\sim 4)}}
\prod_{r=1}^4\left((2l_r+1)^{\frac12}\pn{r}\right) \nn \\
&\times\sum_{lm}(-1)^{-m}(2l+1)
\bep
l_1 & l_4 & l \\
m_1 & m_4 & m
\eep
\bep
l & l_3 & l_2 \\
-m & m_3 & m_2
\eep
\beB
l_1 & l_4 & l \\
L & L & L
\eeB
\beB
l & l_3 & l_2 \\
L & L & L
\eeB, 
\label{mmaction0}
\end{align}
where the summations effectively run in regions constrained 
by the triangular conditions (mentioned in \eqref{triangular}) 
imposed by the $3j$- and $6j$- symbols. 
Hereafter a sum over $l$, $m$ should be understood likewise 
and we will not show their range explicitly.

\subsection{Large-$N$ renormalization group on fuzzy sphere}
In this subsection we demonstrate a new large-$N$ renormalization group (RG)
by taking advantage of the fuzzy sphere. 
As mentioned in the introduction, the main motivation to develop the RG 
on the fuzzy sphere is that there we can introduce the concept of the angular momentum 
in the space of $N\times N$ matrices. This contrasts with the original large-$N$ RG 
proposed in \cite{Brezin:1992yc}, where there is no clear notion of momentum. 
Thus the fuzzy sphere enables us to perform a new RG transformation lowering $N$: $N\rightarrow N-1$ 
by integrating out matrix elements with the maximal angular momentum. 
The difference 
between our RG and that in \cite{Brezin:1992yc} is just the choice of the basis 
in the space of $N\times N$ matrices based on which the large-$N$ RG
is formulated.
There would then be no gauge-invariant notion of matrix elements
of high/low momenta since they are mixed up under $U(N)$ rotations.
However, introducing the kinetic term breaks $U(N)$ and enables us to
define high/low momentum with respect to the choice of the kinetic
term action.
Our formulation then manifestly accords with the spirit of the RG
\cite{Wilson:1973jj}.
The resulting RG is expected to be a local and well-defined transformation and to give rise to a local Lagrangian 
with nice behavior in UV. 
Notice that for the purpose of the large-$N$ RG 
we have to formulate the RG for finite $N$. The fuzzy sphere which can be formulated with finite $N$ 
fits in our purpose, while the noncommutative plane based on the Heisenberg algebra 
does not, at least in a straightforward manner.\footnote
{We will mention a serious problem in use of the fuzzy torus in section \ref{sec:discussions}.} 

Now we formulate our large-$N$ RG. We start from the $N\times N$ matrix model as in \eqref{action}
\begin{align}
S_N=\frac{\rho_N^2}{N}\trn{N}{-\frac{1}{2\rho_N^2}[L_i,\phi_N]^2+\frac{m_N^2}{2}\phi^2+\frac{g_N}{4}\phi^4},
\label{mmaction}
\end{align}
where we have put the subscript $N$ on the parameters in order to keep track of their RG flow.
The RG transformation is carried out through the following two procedures: the first is
coarse-graining,  and the second is rescaling.
For the first, we integrate out modes with the maximal angular momentum, i.e. 
$\plm$ with $l=2L$ and $-2L\leq m\leq 2L$ in the expansion \eqref{phidecomp}. 
Then the number of remaining modes will be $N^2-(4L+1)=(N-1)^2$, which is exactly the number of components 
of $(N-1)\times (N-1)$ matrix. Thus we would reinterpret the resulting action as that of $(N-1)\times (N-1)$ matrix. 
Namely we construct a new map from a matrix model with rank $N$ to that with rank $N-1$. 
Next we make a scale transformation $\rho_N\rightarrow \rho_{N-1}$ in order to recover the original scale, 
and finally read off the change of parameters $m_N^2$ and $g_N$ to $m_{N-1}^2$ and $g_{N-1}$. This is our new RG 
by utilizing the fuzzy sphere.\footnote
{Another RG approach to matrix models based on the configuration space is proposed in \cite{Narayan:2002gv}.} 
In the following, we will present the details of each step in order.

\subsubsection{Coarse-graining}
\label{sec:coarse-graining}

For notational simplicity, let us
divide the space of the angular momentum $\Lambda$ as 
\begin{align}
&\Lambda=\left\{(l,m)\,|\,0\leq l\leq 2L,\,-l\leq m\leq l\right\}\subset\bm Z^2, \nn \\
&\Lo=\left\{(l,m)\,|\,l=2L,\,-2L\leq m\leq 2L\right\}\subset\Lambda, \nn \\
&\Li=\Lambda\setminus\Lo=\left\{(l,m)\,|\,0\leq l\leq2L-1,\,-l \leq m\leq l\right\},
\end{align}
and define correspondingly 
\begin{align}
\plm=\left\{
\begin{array}{ll}
\pon{} & (l,m)\in\Lo \\
\pin{} & (l,m)\in\Li
\end{array}
\right..
\end{align}
Thus we formulate the coarse-graining procedure as 
\begin{align}
S_{N-1}(m_{N-1}^2,g_{N-1})=-\log\int\prod_{m=-2L}^{2L}d\pon{}\,e^{-S_N(m_N^2,g_N)},
\label{RG}
\end{align}
where $S_{N-1}(m_{N-1}^2,g_{N-1})$ is an action of $(N-1)\times (N-1)$ matrix.\footnote
{A similar RG has been proposed in \cite{Vaidya:2001bt} for the purpose of analyzing the noncommutative field theory.
However, it should be noticed that in our large-$N$ RG we construct an $(N-1)\times(N-1)$ matrix model, 
not just integrating the out-modes. As a result, we have to take account of renormalization of noncommutativity 
as discussed in present subsection, which is in contrast to \cite{Vaidya:2001bt}, where the noncommutativity 
is fixed.} 
It is important that our large-$N$ RG formalism respects the rotational symmetry $SO(3)$, 
because it integrates out the whole components of one irreducible representation of $SO(3)$ 
which does not mix modes with different $l$ as in \eqref{rep}. 

In order to make calculation of \eqref{RG} tractable, 
we first note that the kinetic term in \eqref{mmaction} can be decomposed as 
\begin{align}
S_N^{(\text{kin}.)}
&=\frac{\rho_N^2}{N}\trn{N}{-\frac{1}{2\rho_N^2}\left[L_i,\phi\right]^2+\frac12m_N^2\phi^2}=\skinin+\skinout, \nn \\
\skinin&=\sum_{(l,m)\in\Li}\frac12(l(l+1)+\rho_N^2m_N^2)
\phi^{\text{in}\,*}_{lm}\pin{}, \nn \\
\skinout&=\sum_{m=-2L}^{2L}\frac12(N(N-1)+\rho_N^2m_N^2)
\phi^{\text{out}\,*}_{2L\,m}\pon{},
\end{align}
and define 
\begin{align}
Z_0&=\int\prod_{m=-2L}^{2L}d\pon{}\,e^{-\skinout}, \quad 
\vev{\cal O}_0=\frac{1}{Z_0}\int\prod_{m=-2L}^{2L}d\pon{}
{\cal O}\,e^{-\skinout},
\end{align}
then our RG equation \eqref{RG} becomes 
\begin{align}
&S_{N-1}(m_{N-1}^2,g_{N-1})=-\log Z_0+\skinin-\log\vev{e^{-\spot}}_0, \nn \\
&\spot=\frac{\rho_N^2g_N}{4N}\tr_N\phi^4. 
\label{RGbyVEV}
\end{align}
Thus the calculation of $S_{N-1}(m_{N-1}^2,g_{N-1})$ amounts to evaluating 
$\vev{e^{-\spot}}_0$. For this purpose, we classify the potential term 
by the number of out-modes as 
\begin{align}
\spot&=\frac{\rho_N^2g_N}{N}\sum_{i=0}^4V_i, \nn \\
V_0&=\frac14\tr_N\phin^4, \nn \\
V_1&=\trn{N}{\phin^3\phout}=\sum_{m=-2L}^{2L}\pon{}\trn{N}{\phin^3\ton{}}, \nn \\
V_2&=V_2^P+V_2^{NP}, \nn \\
&V_2^P=\trn{N}{\phin^2\phout^2}
      =\sum_{m,m'=-2L}^{2L}\pon{}\pond{}\trn{N}{\phin^2\ton{}\tond{}}, \nn \\
&V_2^{NP}=\frac12\trn{N}{\phin\phout\phin\phout}
         =\frac12\sum_{m,m'=-2L}^{2L}\pon{}\pond{}\trn{N}{\phin\ton{}\phin\tond{}}, \nn \\
V_3&=\trn{N}{\phin\phout^3}
    =\sum_{m_1,m_2,m_3=-2L}^{2L}\pon{1}\pon{2}\pon{3}\trn{N}{\phin\ton{1}\ton{2}\ton{2}}, \nn \\
V_4&=\frac14\tr_N\phout^4
    =\sum_{m_1\sim m_4=-2L}^{2L}\prod_{i=1}^4\pon{1}\trn{N}{\prod_{i=1}^4\ton{i}},
\label{vertices}
\end{align}
where $\phin$ and $\phout$ denote a matrix only with in-modes and out-modes, respectively: 
\begin{align}
\phin=\sum_{(l,m)\in\Li}\pin{}\tlm, \quad 
\phout=\sum_{m=-2L}^{2L}\pon{}\ton{}.
\end{align}
The expectation value can be evaluated by using the propagator of the out-modes 
\begin{align}
\vev{\pon{}\pond{}}_0=\delta_{m+m'}(-1)^mP_N,
\label{propagator}
\end{align}
where $P_N$ does not depend on $m$, $m'$ as 
\begin{align}
P_N=\frac{1}{N(N-1)+\rho_N^2m_N^2}.
\label{propagator2}
\end{align}
By using this, we can perturbatively integrate out $\phout$ as
\begin{align}
-\log\vev{e^{-\spot}}_0=\frac{\rho_N^2g_N}{N}\sum_i\vev{V_i}_0
-\frac12\left(\frac{\rho_N^2g_N}{N}\right)^2\sum_{ij}\vev{V_iV_j}_c+{\cal O}(g_N^3),
\end{align}
where the subscript $c$ means taking the connected part. 
In the rest of the paper, 
we will demonstrate the RG transformation to the second order in $g_N$. 
Noting the $\bm Z_2$-symmetry 
of the original action \eqref{mmaction}, we find that 
\begin{align}
S_{N-1}(m_{N-1}^2,g_{N-1})=&\skinin+\frac{\rho_N^2g_N}{N}\left(V_0+\vev{V_2}_0\right) \nn \\
&-\frac12\left(\frac{\rho_N^2g_N}{N}\right)^2\left(\vev{V_1^2}_c+\vev{V_2^2}_c+\vev{V_3^2}_c
+2\vev{V_1V_3}_c+2\vev{V_2V_4}_c\right) \nn \\
&+{\cal O}(g_N^3)+\text{const.},
\label{RGexp}
\end{align}
where we do not include the expectation values containing only $V_4$,
since they do not depend on $\phin$ and then will not
affect the renormalization of the parameters $m_N^2$, $g_N$.
We postpone carrying out the actual calculations
and first explain the rest of the large-$N$ RG procedure.

\subsubsection{Rescaling}
\label{sec:rescaling}

The second procedure of the RG transformation is the rescaling. 
We will see that it also involves the renormalization of noncommutativity.
After coarse-graining,
we should have a matrix model of size $(N-1)\times (N-1)$, say
$\wphi$, instead of $N\times N$ matrix $\phin$. 
To do it, we relate $\phin$ and $\wphi$ by 
\begin{align}
\frac1N\trn{N}{-\frac12[L_i,\phin]^2}=\frac{1}{N-1}\trn{N-1}{-\frac12[\wL_i,\wphi]^2},
\label{wfrenormalization}
\end{align} 
which implies that 
\begin{align}
\pin{}=\wphi_{lm},
\end{align}
with $\wphi=\sum_{(l,m)\in\Li}\wphi_{lm}\wT_{lm}$, where $\wL_i$ is the $SU(2)$ generator 
\eqref{generator} of spin $L-1/2=(N-2)/2$ representation, 
and $\wT$ is the fuzzy spherical harmonics 
with rank $N-1$. Namely, we fix the wave function renormalization in such a way that 
the kinetic term is canonically normalized. Then it immediately follows that 
\begin{align}
\frac1N\trn{N}{\phin^2}=\frac{1}{N-1}\trn{N-1}{\wphi^2}. 
\label{massrescale}
\end{align}
On the other hand, 
it is nontrivial how the $\trn{N}{\phin^4}$ vertex becomes 
in terms of $\wphi$. For this purpose it is evident that we have to shift all $L$'s 
appearing in the $6j$-symbols in \eqref{mmaction0} by $-1/2$ (recall $L=(N-1)/2$). 
Here the key recursion relation of the $6j$-symbol is \cite{Var}
\begin{align}
&\begin{Bmatrix}
a & b & l \\
L & L & L
\end{Bmatrix}
=\frac{1}{\sqrt{(2L+a+1)(2L-a)(2L+b+1)(2L-b)}} \nn \\
&\times
\Biggl[
-2L\sqrt{(2L+l+1)(2L-l)}
\begin{Bmatrix}
a & b & l \\
L-\frac12 & L-\frac12 & L-\frac12
\end{Bmatrix} 
+\sqrt{(a+1)a(b+1)b}
\begin{Bmatrix}
a & b & l \\
L & L & L-1
\end{Bmatrix}
\Biggr]. 
\label{recursion}
\end{align}
We find that the first term indeed provides the necessary factor for the $\trn{N-1}{\wphi^4}$, 
while the second term does not seem to correspond to natural operation of an 
$(N-1)\times (N-1)$ matrix. 
However, as in the usual RG, we are interested in the low energy regime 
$l_i\ll L$ ($i=1\sim 4$) in $\tr_N(T_{l_1m_1}T_{l_2m_2}T_{l_3m_3}T_{l_4m_4})$ 
and $a$, $b$ in \eqref{recursion} applied to \eqref{mmaction0} 
correspond to low energy modes. 
We then find that the second term is of ${\cal O}(1/N^2)$ compared to the first term, 
which can be therefore neglected in the large-$N$ 
RG discussing corrections of ${\cal O}(1/N)$ when $N\rightarrow N-1$. 
Substituting \eqref{recursion} into an explicit form \eqref{4T2}, a little calculation shows that 
\begin{align}
\trn{N}{\phi^4}=\left(1+\frac1N\right)\trn{N-1}{\wphi^4}+{\cal O}\left(\frac{1}{N^2}\right).
\label{potrescale}
\end{align}

We also rescale $\rho^2_N$ in our RG transformation. 
In the ordinary RG on the two-dimensional square lattice, the lattice spacing 
will be multiplied by $\sqrt{2}$ after the block spin transformation, and 
we rescale the length scale by $1/\sqrt2$ in order to recover the original lattice spacing. 
In the case of the RG of field theory, after integrating out high energy modes 
$\Lambda/b \leq p\leq \Lambda$ (for some $b>1$) with $\Lambda$
the momentum cutoff, we make a 
scale transformation as $p\rightarrow b p$ in order to retrieve the original momentum space. 
Since in the case of the fuzzy sphere 
the maximal momentum is given by the Casimir of the mode with $2L$, we rescale $\rho_N$ as 
\begin{align}
\frac{2L(2L+1)}{\rho_N^2}=\frac{2L(2L-1)}{\rho_{N-1}^2},
\label{fsscale}
\end{align}
namely, in such a way that the maximal momentum will be fixed in a
rotational invariant manner.
Here we define a ratio 
\begin{eqnarray}
b_N^2\equiv \frac{\rho_N^2}{\rho_{N-1}^2}. 
\label{scale}
\end{eqnarray} 
 Thus $b_N^2$ is given by 
\begin{align}
b_N^2=\frac{N}{N-2}, 
\label{fsb}
\end{align}
by which we make a scale transformation in order to recover the original scale 
as in the usual RG of field theory. Notice that the degrees of freedom in a unit volume 
would be then multiplied by $N/(N-1)$, which is consistent with the picture that 
the fuzzy sphere defined by $N\times N$ matrices  
describes $N$ quanta with volume given by $\alpha_N^2$ as in \eqref{alphadef}:\footnote
{For more details, see e.g. \cite{Kawai:2003yf}.} 
\begin{align}
\alpha_N^2=\frac{4}{N^2-1}\rho_N^2.
\end{align}
Correspondingly, it should be noticed that under the scale transformation \eqref{fsscale},  
$\alpha_N$ is invariant: 
\begin{align}
\frac{\alpha_N^2}{\alpha_{N-1}^2}=1+{\cal O}\left(\frac{1}{N^2}\right). 
\end{align}
In other words, in the case of fuzzy sphere we take the large-$N$ limit 
with the physical length scale $\alpha^2$ fixed, 
as the continuum limit in the lattice QCD with the physical pion mass fixed.   
Since $\alpha$ is the physical scale characterizing the fuzzy sphere, this limit 
is natural, and accordingly we examine it via the large-$N$ RG which respects $\alpha$.

On the other hand, it is known that by taking another large-$N$ limit, 
we can obtain the noncommutative field theory (NCFT) \cite{Minwalla:1999px, Szabo:2001kg} 
on the flat two-dimensional plane (see e.g. \cite{Iso:2001mg, Chu:2001xi}). 
In this case, the large-$N$ limit is taken with the noncommutativity $\theta$ in the NCFT fixed as 
\cite{Chu:2001xi} 
\begin{align}
N\rightarrow\infty \quad \text{with} \quad \theta=\frac{2\rho^2}{N}:~\text{fixed}. 
\label{NCFTlimit}
\end{align}
Hence in order to examine the NCFT via our large-$N$ RG, we should fix
$\theta_N$.
This implies that 
\begin{align}
\frac{\rho_N^2}{N}=\frac{\theta_N}{2}=\frac{\theta_{N-1}}{2}=\frac{\rho_{N-1}^2}{N-1}.
\label{thetadef}
\end{align}
Namely, in this case we have instead of \eqref{fsb}
\begin{align}
b_N^2=\frac{N}{N-1}.
\label{NCb}
\end{align}

\subsubsection{Change of the parameters}

After evaluating (\ref{RGexp}) and performing the scale transformation,
the renormalized action can schematically be written as
\begin{align}
S_{N-1}
=\frac{\rho_{N-1}^2}{N-1}\tr_{N-1}
\Biggl[&-\frac{1}{2\rho_{N-1}^2}\biggl(1+K_N(N,m^2_N,g_N)\biggr) [\wL_i,\wphi]^2
+\frac{b_N^2}{2}\biggl(m_N^2+M_N(N,m^2_N, g_N)\biggr)\wphi^2\nn \\
&+\frac{b_N^2}{4}\biggl(g_N+G_N(N,m^2_N,g_N)\biggr)\wphi^4 \Biggr] \nn\\
+(\mbox{others}),
\label{RGfinal}
\end{align}
where we have taken account of \eqref{massrescale}, \eqref{potrescale} and \eqref{scale}. 
In general, $K(N,m^2_N,g_N)\neq 0$,
and a further rescaling of $\wphi$ has to be done in order to make the
kinetic term canonical. 
The resulting action should be compared with
\begin{align}
    S_{N-1} =& \frac{\rho_{N-1}^2}{N-1} \tr_{N-1}
\left[
\frac{1}{2 \rho_{N-1}^2} [\wL_i,\wphi]^2
+ \frac{m_{N-1}^2}{2} \wphi^2
+ \frac{g_{N-1}}{4} \wphi^4
\right] \,.
\end{align}
Now we read off the change of parameters as
\begin{align}
&m_{N-1}^2=b_N^2\frac{m_N^2+M_N(N,m^2_N,g_N)}{1+K_N(N,m^2_N,g_N)}, \nn \\
&g_{N-1}=b_N^2\frac{g_N+G_N(N,m_N^2,g_N)}{ (1+K_N(N,m^2_N, g_N) )^2}.
\label{RGE}
\end{align}
These are RG flow equations for $m^2$ and $g$, as $N\rightarrow N-1$.  

In the last line of (\ref{RGfinal}),
$(\text{others})$ includes terms of the form which did not appear
in the original action, induced through the coarse-graining procedure.
It contains higher order interaction terms $\tr_{N-1}\,
\tilde\phi^{2n}$
 ($n\ge 3$) 
and multi-trace deformations.
They would induce the RG flow in an enlarged space of coupling constants.
However, to the order of the perturbation in the present paper, they
are not relevant.
It also contains various ``derivative expansions'' of kinetic and
potential terms, and highly ``nonlocal" terms which are related to
nonplanar diagrams in perturbation theory. 
In the following subsections, we will see how they appear and how to
deal with them.
\\

In following subsections we will concretely evaluate
the terms in (\ref{RGexp}) and determine the corrections shown in (\ref{RGfinal}).
There we will find several quite interesting and characteristic aspects 
of our large-$N$ RG, namely, appearance of ``antipode'' transformed
fields and multi-trace operators,  
and ``derivative'' expansions in the space of matrices. 

\subsection{${\cal O}(g_N)$ and appearance of antipode transformation}
\label{orderg}
\begin{figure}
\begin{center}
\subfigure[][]{%
	\includegraphics[scale=.5,bb=100 100 400 300]{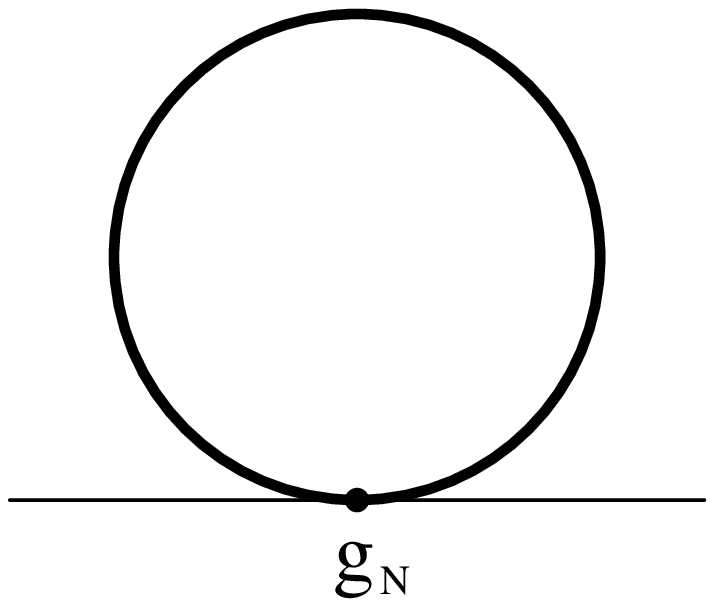}
	\label{fig:deltamp}
}
\hskip 5em
\subfigure[][]{%
	\includegraphics[scale=.5,bb=150 100 400 300]{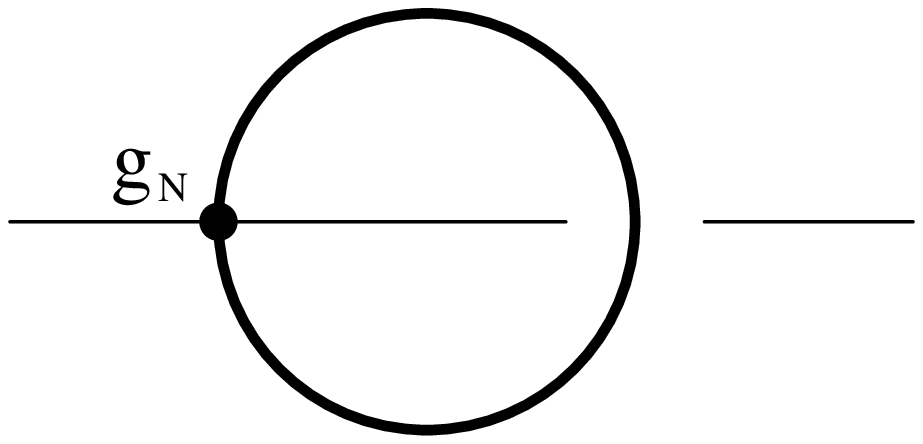}
	\label{fig:deltamnp}
}
\end{center}
\caption{%
The planar \subref{fig:deltamp} and nonplanar \subref{fig:deltamnp} contribution 
of ${\cal O}(g_N)$. The bold line represents the propagator of the out-mode \eqref{propagator}.}
\label{Og}
\end{figure}

As a demonstration let us calculate contributions of ${\cal O}(g_N)$ in \eqref{RGexp}. 
{}From \eqref{vertices} and \eqref{propagator}, we obtain 
\begin{align}
\vev{V_2^P}_0&=\sum_{m,m'}\vev{\pon{}\pond{}}_0\trn{N}{\phin^2\ton{}\tond{}} \nn \\
&=P_N\sum_m(-1)^m\trn{N}{\phin^2\ton{}\tom{}}.
\end{align}
Here we note the formula given in \eqref{coro} in appendix \ref{app:tlm}
\begin{align}
\sum_{m=-2L}^{2L}(-1)^m\ton{}\tom{}
=(2N-1)\bm 1_{N}, 
\end{align} 
and get 
\begin{align}
\vev{V_2^P}_0=(2N-1)P_N\tr_N\phin^2,
\label{deltamp}
\end{align}
which therefore contributes to the mass correction. This contribution comes from a planar diagram 
given in \subref{fig:deltamp} in Fig.\ref{Og}.

On the other hand, we also have a contribution from the nonplanar diagram shown in \subref{fig:deltamnp}
in Fig.\ref{Og}
\begin{align}
\vev{V_2^{NP}}_0&=\frac12\sum_{m,m'}\vev{\pon{}\pond{}}_0\trn{N}{\phin\ton{}\phin\tond{}} \nn \\
&=\frac{P_N}{2}\sum_m(-1)^m\trn{N}{\phin\ton{}\phin\tom{}}.
\label{V2NPtemp}
\end{align}
Hence the nonplanar diagram generates a new operator. 
In order to elaborate on it, we apply the formula in \eqref{similarity} in appendix \ref{app:tlm}
as 
\begin{align}
\sum_m(-1)^m\trn{N}{\phin\ton{}\phin\tom{}}
=N(2N-1)
\sum_{(l,m)\in\Li}
\beB
L & L & l \\
L & L & 2L
\eeB
\pin{}\trn{N}{\phin(-1)^l\tlm}.
\end{align}
It is then natural to define an operation on $N\times N$ matrices as 
\begin{align}
\tlm \quad \longmapsto \quad \tlm^A\equiv (-1)^l\tlm,
\end{align}
thus  
\begin{align}
\phin=\sum_{(l,m)\in\Li}\pin{}\tlm \quad \longmapsto \quad \phinA=\sum_{(l,m)\in\Li}(-1)^l\pin{}\tlm.
\label{antipode}
\end{align}
Although this transformation looks strange in the space of matrices, it has a natural interpretation 
on the fuzzy sphere. In fact, it corresponds to a discrete transformation on $S^2$
\begin{align}
\ylm(\theta,\varphi) \quad \longmapsto \quad (-1)^l\ylm(\theta,\varphi)=\ylm(\pi-\theta,\varphi+\pi),
\end{align}
namely, it transports a field to the antipode point on $S^2$. \eqref{antipode} 
is therefore a natural counterpart on the fuzzy sphere of this transformation and hereafter 
we call it antipode transformation. Using this, \eqref{V2NPtemp} becomes 
\begin{align}
\vev{V_2^{NP}}_0
=\frac{P_N}{2}N(2N-1)
\sum_{(l,m)\in\Li}
\beB
L & L & l \\
L & L & 2L
\eeB
\pin{}\trn{N}{\phin\tlm^A}.
\label{V2NPtemp2}
\end{align}
Furthermore, we can make a sort of ``derivative expansion" of this expression. 
In fact, from \eqref{p308(16)} the $6j$-symbol in \eqref{V2NPtemp2} has a large-$L$ (thus large-$N$) expansion\footnote{%
Throughout this paper, we concentrate on the case with the ``in''
fields carrying low enough momenta.
Therefore in the expression of the perturbation theory, we write
errors simply as $\mathcal{O}(1/L^2)$ or $\mathcal{O}(1/N^2)$,
though it could also depend on the momenta of
associated ``in'' fields.} 
\begin{align}
\beB
L & L & l \\
L & L & 2L
\eeB
&\simeq \frac{(-1)^l}{2L+1}P_l\left(-1+\frac{1}{L+1}\right) \nn \\
&=\frac{1}{2L+1}\left(1-\frac{l(l+1)}{2(L+1)}+{\cal O}\left(\frac{1}{L^2}\right)\right),
\label{6jexp1}
\end{align}
to give
\begin{align}
\vev{V_2^{NP}}_0
=\frac{P_N}{2}(2N-1)
\left[\trn{N}{\phin\phinA}-\frac{1}{N}\trn{N}{\phin\Lap{\phinA}}+{\cal O}\left(\frac{1}{N^2}\right)
\right],
\label{V2NPtemp3}
\end{align}
where we recall $L=(N-1)/2$ and \eqref{Casimir}. It is quite interesting that 
owing to the fuzzy sphere structure, the space of $N\times N$ matrices admits 
the derivative expansion of an operator. Here higher derivative terms come from 
the $l$-dependent terms of the expansion of the Legendre polynomial in \eqref{6jexp1} 
in terms of $1/L$ and are suppressed by $1/N$ as such. 
Notice that this expansion should be 
in terms of the Laplacian \eqref{Casimir}, because the fuzzy sphere preserves 
the rotational symmetry $SO(3)$, which is respected in our large-$N$ RG formalism.  
Thus we find that the operator on the right-hand side in \eqref{V2NPtemp2}
contains $\trn{N}{\phin\phinA}$ and its derivatives\footnote{%
At a glance, the antipode transformation seems to correspond
to exchange of two matrices.
However, as shown in
appendix \ref{app:antipode}, it is related to  the reverse of
the ordering of all matrices inside a trace.}.
We point out here that 
new operators that are not contained in the original Lagrangian \eqref{mmaction} 
are thus generated in our RG, but that they may be regarded as small corrections. 
For example, the first term in \eqref{V2NPtemp3} agrees with the overlap 
between a wave function and its antipode 
\begin{align}
\frac1N\trn{N}{\phin\phinA}
=\int\frac{d\Omega}{4\pi}\phi^{\text{in}}(\theta,\varphi)\phi^{\text{in}}(\pi-\theta,\varphi+\pi)
=\sum_{(l,m)\in\Li}(-1)^l|\pin{}|^2,
\end{align}
which is always smaller than $\frac1N\trn{N}{\phin^2}=\sum_{(l,m)\in\Li}|\pin{}|^2$ in \eqref{deltamp} 
coming from the planar diagram. In fact, it is easy to check that for the Gaussian wave function 
(under the stereographic projection) the former is exponentially smaller than the latter with respect to 
the inverse square of the width. However, since in the RG we are interested in low energy modes, 
it is not obvious that the overlap is small enough to ignore. Thus in the fixed point analysis 
given in section \ref{sec:fp} we will concentrate on the flows
 of $m_N^2$ and $g_N$ up to ${\cal O}(g_N^2)$ 
and simply assume that to this order, interactions involving $\phin^A$ would not affect them considerably, 
namely, would not modify \eqref{RGE} so much.\footnote
{Here we should notice that 
it is likely that an interaction with $\phin^A$ originating from a nonplanar diagram 
is not irrelevant in itself near a fixed point. 
In fact, it is shown in \cite{Chu:2001xi} that a contribution from the 1PI nonplanar two-point function 
$\Gamma^{(2)}_{\text{nonplanar}}$ (corresponding to $\vev{V_2^{NP}}$ with all modes running the loop integrated) 
gives rise to the IR singularity in the large-$N$ NCFT limit \eqref{NCFTlimit} 
and hence becomes the origin of the UV/IR mixing. 
Since the large-$N$ limit should be captured by a fixed point in the large-$N$ RG, 
this suggests that if an interaction involving $\phin^A$ has something to do with $\Gamma^{(2)}_{\text{nonplanar}}$ 
which becomes quite large in the IR in the large-$N$ limit, 
it would be rather relevant around a corresponding fixed point.  
We will further make a few comments on such an interesting aspect in section \ref{sec:discussions}.} 
Here it is important to note that as we see by comparing the coefficients in 
\eqref{deltamp} and \eqref{V2NPtemp3}, the nonplanar contribution is not suppressed 
by ${\cal O}(1/N)$ in itself compared to the planar one, but the momentum dependence 
(in the present case $(-1)^l$ in the antipode transformation) can make it suppressed. 
Recall that this is also the case with the NCFT, where nonplanar diagrams are in general dropped 
in a high energy regime by the oscillating phase depending on momenta.   

\subsection{${\cal O}(g_N^2)$ and appearance of double trace operator}
\label{sec:Og2}

\subsubsection{Vertex correction}

Among contributions of ${\cal O}(g_N^2)$ shown in the second line in \eqref{RGexp}, 
$\phi^4$ vertex correction originates only from $\vev{V_2^2}_c$. 
Thus in this subsection,
we focus on this part and present only the results of the calculation
of the other contributions. 
The details are given in appendix \ref{app:Og2}. 

\begin{figure}
\begin{center}
\subfigure[][]{%
	\includegraphics[scale=.5,bb=100 100 400 300]{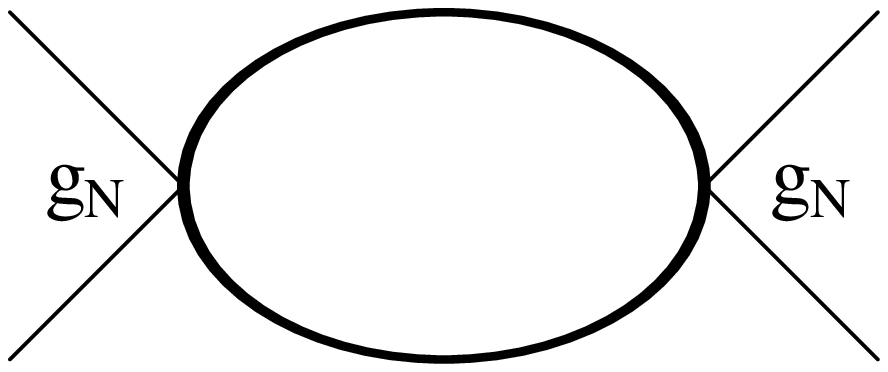}
	\label{fig:V2P2p}
}
\hskip 5em
\subfigure[][]{%
	\includegraphics[scale=.5,bb=150 100 400 300]{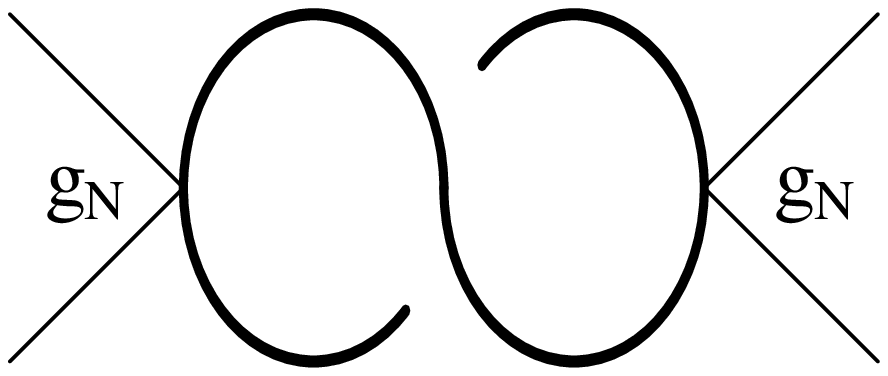}
	\label{fig:V2P2np}
}
\end{center}
\caption{%
The planar \subref{fig:deltamp} and nonplanar \subref{fig:deltamnp} contribution 
in $\vev{{V_2^P}^2}_c$. The bold line represents the propagator of the out-mode \eqref{propagator}.}
\label{V2P2}
\end{figure}

Since $V_2=V_2^P+V_2^{NP}$ as in \eqref{vertices}, let us begin with $\vev{{V_2^P}^2}_c$. From \eqref{vertices},   
\begin{align}
\vev{{V_2^P}^2}_c&=\sum_{m_1,m_2, m'_1,m'_2}
\left(\vev{\pon{1}\pond{2}}_0\vev{\pon{2}\pond{1}}_0+\vev{\pon{1}\pond{1}}_0\vev{\pon{2}\pond{2}}_0\right) \nn \\
&\hspace{2cm}\times\trn{N}{\phin^2\ton{1}\ton{2}}\trn{N}{\phin^2\tond{1}\tond{2}} \nn \\
&=P_N^2\sum_{m_1,m_2}(-1)^{m_1+m_2}\biggl(\trn{N}{\phin^2\ton{1}\ton{2}}\trn{N}{\phin^2\tom{2}\tom{1}} \nn \\
&\hspace{3.8cm}+\trn{N}{\phin^2\ton{1}\ton{2}}\trn{N}{\phin^2\tom{1}\tom{2}}\biggr).
\label{V2P2temp}
\end{align}
The first and second term correspond to the planar and nonplanar diagram shown in Fig.\ref{V2P2} 
\subref{fig:V2P2p} and \subref{fig:V2P2np}, respectively. 
Important observation here is that double trace terms as above are in general generated in the large-$N$ RG. 
At first sight, they seem to be new interactions arising in our RG and 
do not play any role in renormalization of parameters $\rho_N$, $m_N^2$ and $g_N$ 
in the original action \eqref{mmaction}. But this is not the case. 
Recall that in the usual RG of the $\phi^4$ scalar field theory we also get 
(in the configuration space) from the diagram in \subref{fig:V2P2p} in Fig. \ref{V2P2}
\begin{align}
\frac{g^2}{8}\int dxdy \Delta(x-y)^2\,\phi(x)^2\phi(y)^2,
\end{align}  
where $\Delta(x-y)=\vev{\phi(x)\phi(y)}$ is the propagator of the scalar field. 
Namely, bi-local interactions are generated in general in the RG corresponding to the double trace 
term in \eqref{V2P2temp}. However, we know that if the propagator $\Delta(x-y)$ is 
for a highly massive mode of $\phi$, it would rapidly damp as $x$ and $y$ are separated. 
In such a case the derivative expansion as 
\begin{align}
\frac{g^2}{8}\int dx\left[\left(\int dz \Delta(z)^2\right)\phi(x)^4
+\left(\int dz \frac{z^2}{2} \Delta(z)^2\right)
\phi(x)^2\partial_x^2\phi(x)^2+\cdots\right],
\label{bilocalexp}
\end{align}
is valid at least for a low energy mode of $\phi$ of interest in the RG. 
Namely, bi-local interactions generated by the RG can be decomposed as a sum of local interactions 
and there higher derivative interactions are suppressed if we integrated out sufficiently high energy modes. 
In other words, the RG by integrating only high energy modes only generates local interactions. 
Notice that the first term in \eqref{bilocalexp} contributes to $\phi^4$ vertex correction. 

Hence in our large-$N$ RG we also have to make a ``derivative" expansion as in \eqref{bilocalexp} 
in the space of matrices and to read off what kind of ``local" interactions arise 
as a result of the RG. Interestingly enough, it is indeed possible in our formalism 
because our RG is based on the fuzzy sphere structure and we in fact integrate out 
modes with the maximal momentum. 
As illustration let us consider the planar contribution 
of $\vev{{V_2^{P}}^2}_c$ given by the first term in \eqref{V2P2temp}. By using the fusion coefficient 
\eqref{fuzzyfusion} with \eqref{fusion} in appendix \ref{app:tlm}, 
\begin{align}
\left.\vev{{V_2^P}^2}_c\right|_{\text{planar}}=&P_N^2(4L+1)^2N\sum_{m_1,m_2}(-1)^{m_1+m_2}
\sum_{(l,m),(l',m')}\sqrt{(2l+1)(2l'+1)}(-1)^{l+m+l'+m'} \nn \\
&\times
\bep
2L & 2L & l \\
m_1 & m_2 & -m
\eep
\bep
2L & 2L & l' \\
-m_2 & -m_1 & -m'
\eep
\beB
2L & 2L & l \\
L & L & L
\eeB 
\beB
2L & 2L & l' \\
L & L & L
\eeB \nn \\
&\times\trn{N}{\phin^2\tlm}\trn{N}{\phin^2\tnd{}}.
\label{V2P2pstart}
\end{align}
Then we can take the summation over $m_1$, $m_2$ first by the formula \eqref{p453(3)} 
in appendix \ref{app:3nj} to get 
\begin{align}
\left.\vev{{V_2^P}^2}_c\right|_{\text{planar}}=&P_N^2(2N-1)^2N
\sum_{lm}
\beB
2L & 2L & l \\
L & L & L
\eeB^2 
\trn{N}{\phin^2\tlm}\trn{N}{\phin^2\tlm^\dagger{}}.
\label{V2P2ptemp}
\end{align}
Now we make a ``derivative" expansion. In the large-$N$ RG we are interested in 
sufficiently low energy modes $\pin{}$ with $l\ll L$ as in the usual RG in field theory, 
and then \eqref{Tortho} tells us that the internal momentum $l$ in \eqref{V2P2ptemp} 
should be also sufficiently smaller than $L$ 
so that the traces will not vanish. 
Thus as far as $\phin$ only with low energy modes is concerned, 
we can evaluate the $6j$-symbol in  \eqref{V2P2ptemp} under $l \ll N$ 
by using \eqref{p308(16)} as 
\begin{align}
\beB
2L & 2L & l \\
L & L & L
\eeB 
&\simeq
\frac{(-1)^l}{\sqrt{(4L+1)(2L+1)}}P_l\left(\sqrt{\frac{2L+1}{2L+2}}\right) \nn \\
&=\frac{(-1)^l}{\sqrt{(2N-1)N}}\left(1-\frac{l(l+1)}{4N}+{\cal O}\left(\frac{1}{N^2}\right)\right).
\label{6jexp}
\end{align}
Plugging this into \eqref{V2P2ptemp}, we obtain
\begin{align}
\left.\vev{{V_2^P}^2}_c\right|_{\text{planar}}=P_N^2(2N-1)
\sum_{lm}
\biggl(
&\trn{N}{\phin^2\tlm}\trn{N}{\phin^2\tlm^\dagger{}} \nn \\
&-\frac{1}{2 N}\trn{N}{\phin^2\Lap{\tlm}}\trn{N}{\phin^2\tlm^\dagger} \nn \\
&+{\cal O}\left(\frac{1}{N^2}\right)\biggr).
\label{V2P2pexp}
\end{align}
Then we see that the first term in fact becomes the single trace by virtue of 
the completeness of $\tlm$ \eqref{merging}. It is rather technical but interesting 
that in the second term we can also take the summation over $l$, $m$ by applying \eqref{merging}, 
because by the ``partial integration" 
\begin{align}
\trn{N}{\phin^2\Lap{\tlm}}\trn{N}{\phin^2\tlm^\dagger}=\trn{N}{\Lap{\phin^2}\tlm}\trn{N}{\phin^2\tlm^\dagger}.
\end{align}
As a consequence we have 
\begin{align}
&\left.\vev{{V_2^P}^2}_c\right|_{\text{planar}} \nn \\
&=P_N^2(2N-1)N
\left[\trn{N}{\phin^4}-\frac{1}{2N}\trn{N}{\Lap{\phin^2}\phin^2}+{\cal O}\left(\frac{1}{N^2}\right)\right].
\label{V2P2p}
\end{align}
Namely, we have the vertex correction of ${\cal O}(g_N^2)$ from the planar diagram. 
Detailed calculations presented in appendix \ref{app:Og2} show that 
the other contributions in $\vev{V_2^2}_c$ always generate $\phi^4$ vertex corrections 
with both $\phin$ and $\phin^A$:  
\begin{align}
&\left.\vev{{V_2^P}^2}_c\right|_{\text{nonplanar}} \nn \\
&=P_N^2(2N-1)N
\left[\trn{N}{\phin^2{\phinA}^2}-\frac{1}{2N}\trn{N}{\Lap{\phin^2}{\phinA}^2}+{\cal
    O}\left(\frac{1}{N^2}\right)\right] , \nn \\
&\vev{V_2^PV_2^{NP}}_c \nn \\
&=P_N^2(2N-1)N
\biggl[\trn{N}{\phin^3\phinA}
-\frac{1}{2N}\trn{N}{\Lap{\phin}\phin^2\phinA} \nn \\
&\hspace{5.9cm}-\frac{1}{2N}\trn{N}{\Lap{\phinA}\phin^3}+{\cal O}\left(\frac{1}{N^2}\right)\biggr], \nn \\
&\vev{{V_2^{NP}}^2}_c \nn \\
&=\frac12P_N^2(2N-1)N
\biggl[\trn{N}{\phin\phinA\phin\phinA}
-\frac{2}{N}\trn{N}{\Lap{\phin}\phinA\phin\phinA}
\nn\\&\hskip5.9cm
-\frac{1}{2N}\trn{N}{\Lap{\phin \phinA }\phin\phinA}
+{\cal O}\left(\frac{1}{N^2}\right)\biggr],
\label{Vnpcorrections}
\end{align}
where we notice that all derivative corrections come in as the Laplacian which reflects 
the fact that our large-$N$ RG as well as the fuzzy sphere respects the $SO(3)$ symmetry, 
as anticipated below \eqref{V2NPtemp3}.  
Therefore the RG again generates new interactions containing both $\phin$ and $\phin^A$ 
that are not present in the original action. For the same reason as in \eqref{V2NPtemp3} 
we regard them as small corrections  and negligible in discussion of the RG of the parameters 
$\rho_N^2$, $m_N^2$ and $g_N$ in the original Lagrangian \eqref{mmaction}.

\subsubsection{Mass correction}

Let us examine the other terms than $\vev{V_2^2}_c$ in the second line in \eqref{RGexp}. 
It is obvious that they contribute to correction in the quadratic order of $\phin$. 
The results are summarized as: 
\begin{enumerate}
\item $\vev{V_1^2}_c$ vanishes in sufficiently low energy regime where the momentum $l$ of $\phin$ 
is much smaller than the cutoff $2L$. 
\item $\vev{V_3^2}_c$ is exponentially suppressed by $L$ when $L\gg 1$.
\item $\vev{V_1V_3}_c=0$ by the momentum conservation.
\item $\vev{V_2^P V_4}_c\simeq P_N^32(2N-1)^2N\trn{N}{\phin^2},$ which arises from the planar diagram.
\item $\vev{V_2^{NP} V_4}_c$ again provides derivative
expansions of terms with both the usual and the antipode fields.
\end{enumerate}
Therefore only $\vev{V_2^P V_4}_c$ part contributes to the mass correction. 
The details of the calculations are shown in appendix \ref{app:Og2}.

\subsection{Flow equations}
\label{subsec:floweq}
So far we see that when $\phin$ is in a sufficiently low energy regime and $L \gg 1$, 
\eqref{RGexp} becomes  
\begin{align}
S_{N-1}=
\tr_{N}\Biggl[&-\frac{1}{2N}[L_i,\phin]^2+\frac{\rho_N^2m_N^2}{2N} \phin^2
+\frac{\rho_N^2g_N}{N}\left(\frac14\phin^4+P_N(2N-1)\phin^2\right) \nn \\
&-\frac12\left(\frac{\rho_N^2g_N}{N}\right)^2\left(P_N^2(2N-1)N\phin^4+4P_N^3(2N-1)^2N\phin^2\right)\Biggr]
+\cdots
 \nn \\
=\frac{\rho_N^2}{N}\tr_N\Biggl[&-\frac{1}{2\rho_N^2}[L_i,\phin]^2 
+\frac12\biggl(m_N^2+g_NB_1(N,m_N^2)-\rho_N^2g_N^2B_1(N,m_N^2)B_2(N,m_N^2)\biggr)\phin^2\nn \\
&+\frac14\left(g_N-\rho_N^2g_N^2B_2(N,m_N^2)\right)\phin^4\Biggr]
+\cdots \,,
\label{RGbyN}
\end{align}
where
\begin{align}
&B_1(N,m_N^2)=2(2N-1)P_N, \nn \\
&B_2(N,m_N^2)=2(2N-1)P_N^2,
\end{align}
with $P_N$ given in \eqref{propagator2}. 
In the above, $\cdots$ denotes the terms including 
$\phin$ independent constants, 
$\mathcal{O}(g_N^3)$, antipode
fields, and $1/N$ suppressed derivative expansions.
They are all omitted in the following RG analysis.
Now the RG flow equations (\ref{RGE}) become
\begin{align}
&m^2_{N-1}=b_N^2\left(m_N^2+g_NB_1(N,m_N^2)-\rho_N^2g_N^2B_1(N,m_N^2)B_2(N,m_N^2)\right), \nonumber \\
&g_{N-1}=b_N^2(g_N-\rho_N^2g_N^2B_2(N,m_N^2)),
\label{RGE2}
\end{align} 
with keeping ${\cal O}(g_N^2)$ terms.

In the next section we analyze the RG transformation \eqref{RGE2}. 
Before closing this section we comment on a couple of 
features of our large-$N$ RG equation in order. 
\begin{enumerate}
\item The mass corrections of ${\cal O}(g_N)$, ${\cal O}(g_N^2)$ 
and the vertex correction of ${\cal O}(g_N^2)$ arise from the planar diagrams 
in $\vev{V_2^P}_0$, $\vev{V_2V_4}_c$, and $\vev{{V_2^P}^2}_c$, respectively. 
Nonplanar diagrams generate terms with the antipode fields, which are
neglected in the current analysis.
\item No kinetic term correction appears.
This reflects the fact that the result of $\vev{V_2^P}_0$ 
in \eqref{deltamp} is exact and does not have any derivative corrections.  
\item Since our large-$N$ RG is constructed by integrating out the modes 
with the highest energy, it should be in a sense local as in the case of the usual RG. 
In fact, all corrections carry the propagator $P_N$ given in \eqref{propagator}, 
which is highly suppressed for large-$N$. Furthermore, our RG admits the ``derivative" 
expansion as mentioned in section \ref{sec:Og2} by which the double trace corrections 
can be written as a sum of the single traces. This fact also reflects locality 
of our RG, because the reason why we can utilize the asymptotic formula of $6j$-symbol 
as in \eqref{6jexp} is that we have integrated the modes with $l=2L$ and hence 
its upper entries become the cutoff $2L$ itself. Accordingly, we have observed that 
the ``derivative'' corrections are suppressed by $1/N$ as in \eqref{V2P2p}.
\end{enumerate}

\section{Fixed point analysis}
\label{sec:fp}
In this section we examine Gaussian and nontrivial fixed points of our RG transformation \eqref{RGE2} 
and discuss their properties. We also determine the 
scaling dimensions of small perturbations around these fixed points
from \eqref{RGE2}. 
First we discuss fixed points 
corresponding to field theories, and then turn to the two-dimensional gravity 
originally considered in \cite{Brezin:1992yc}. 
Results in this section are compared to preceding works
\cite{Martin:2004un}\cite{Panero:2006bx}\cite{Das:2007gm}\cite{Steinacker:2005wj},
regarding nonperturbative studies of $\phi^4$ theory on the fuzzy sphere via the matrix model 
by different methods.

\subsection{Gaussian fixed point}
It is evident that $m_*=g_*=0$ is indeed the fixed point (Gaussian
fixed point) of the RG transformation \eqref{RGE2}.
Near the Gaussian fixed point,
we linearize \eqref{RGE2} with respect to small perturbations $m^2 \ll 1$ 
and $g \ll 1$ as
\begin{align}
&m_{N-1}^2
=b_N^2\left(m_N^2 + g_NB_1(N)\right), \nn \\
&g_{N-1}=b_N^2g_N
\label{oRGE}.
\end{align}
Here we expanded  $B_1(N, m_N^2)$ and $B_2(N, m_N^2)$ 
as
\begin{align}
    B_1(N, m_N^2) =& B_1(N) + B_{1,1}(N) m_N^2 + \cdots \,,
\quad
    B_2(N, m_N^2) =& B_2(N) + B_{2,1}(N) m_N^2 + \cdots \,,
\end{align}
where
\begin{align}
B_1(N)=\frac{2(2N-1)}{N(N-1)} \,,
\qquad
B_2(N) = \frac{2(2N-1)}{N^2 (N-1)^2} \,.
\end{align}
In order to diagonalize the RG transformation 
near the Gaussian fixed point, we assume the eigenvector as
$m^2_N-\bar{m}^2(g_N)$, where  $\bar{m}^2(g_N)=A_Ng_N+B_Ng_N^2+\cdots$.
We then need to solve\footnote{%
As the readers can see, 
the coefficient $A_N$ in $\bar{m}^2(g)$ depends on the step of RG
transformation, namely varies as $N \rightarrow N-n$. Therefore
$\bar{m}^2(g)$ is slightly different from the critical line of the RG
flow. The leading behavior of $A_N$ shown here
approximates the slope of critical line in the vicinity of the fixed point.}
\begin{align}
m_{N-1}^2-\bar{m}^2(g_{N-1})&=b_N^2\left(m_N^2-\bar{m}^2( g_N)\right), \nn \\
g_{N-1}&=b_N^2g_{N} \,.
\label{linearRGE}
\end{align}
These equations determine $\bar{m}^2(g_N)$ as
\begin{align}
\bar{m}^2(g_N)= -\left(4\log N\right)g_N+O\left(\frac{1}{N}\right).
\end{align}
Thus $\bar{m}^2$ approximately give a critical line of the RG flow, and then we find that 
the critical line is vertical in $m^2-g$ plane in large-$N$ limit. 
This feature originates 
from the fact that $m_N^2$ and $g_N$ have the same scaling dimension and thus 
it is peculiar to two dimensions. In fact, this is also the case with 
the RG of the ordinary two-dimensional field theory where the critical
line becomes vertical. This vertical line can be derived in a
different manner. 
In order to regularize the slope of the vertical line, we introduce an extra 
tiny $N$-dependence $c(N)\sim N^{-\varepsilon}$ to $g_N$ in \eqref{mmaction} 
\begin{align}
S_N=\frac{\rho_N^2}{N}\trn{N}{-\frac{1}{2\rho_N^2}[L_i,\phi_N]^2+\frac{m_N^2}{2}\phi^2
+\frac{c(N)g_N}{4}\phi^4},
\label{rmmaction}
\end{align}
and repeat our large-$N$ RG to obtain the RG equation 
\begin{align}
&m_{N-1}^2=b_N^2\left(m_N^2+c(N)g_NB_1(N,m^2_N)-c(N)^2\rho_N^2 g_N^2 B_1(N,m^2_N) B_2(N,m^2_N) \right), 
\label{rmRGE}\\
&g_{N-1}=\frac{c(N)}{c(N-1)}b_N^2\left(g_N-\rho_N^2c(N)g_N^2B_2(N, m^2_N)\right).
\label{rgRGE}
\end{align}
Now critical line $m^2_c(g)$ near the Gaussian fixed point can be determined as  
\begin{align}
m_c^2(g)=-\frac{1}{\varepsilon}\frac{2(2N-1)}{N-1}g\rightarrow -\frac{4}{\varepsilon}g
~~~(N\rightarrow\infty).
\end{align}
Thus we also find the vertical critical line. Compared to the case of the ordinary 
field theory, we see that the introduction of $c(N)\sim
N^{-\varepsilon}$ is an analog of considering
 $(2+\varepsilon)$--dimension. Furthermore, \eqref{rmRGE} and \eqref{rgRGE} tell us that 
the eigenvalues of the RG transformations are  $b_N^2$ for $m_N^2$ and $\frac{c(N)}{c(N-1)}b_N^2$ for $g_N$. 
Now we propose how 
we read off the scaling dimension in our large-$N$ RG. Recall that we make 
the scale transformation by $b_N^2=\rho_N^2/\rho_{N-1}^2$ in order to compensate 
the maximal momentum integrated out in the large-$N$ RG. Therefore, the scaling dimension 
should be given by how many powers of $b_N$ an eigenvalue of the RG transformation has, 
as in the standard way to extract the scaling dimension in the usual RG. 
In other words, since $b_N=1+{\cal O}(1/N)$, we can read off 
the scaling dimension by observing deviation of ${\cal O}(1/N)$ from 1 
in an eigenvalue of the linearized RG transformation.\footnote
{Since we have just changed $N$ by one in the large-$N$ RG, the eigenvalue must be 1 in the leading order 
in the $1/N$ expansion.} Applying this method to \eqref{rmRGE} and \eqref{rgRGE}, 
we conclude that our Gaussian fixed point is an UV fixed point and the scaling dimensions of $m^2$ and $g$ are both 2 for infinitesimal $\varepsilon$. 
It is in fact consistent with the result from the mean-field, or the Landau theory, 
as it should be near the Gaussian fixed point.  

The critical line found in present analysis also have been seen in
numerical works
\cite{Martin:2004un, Panero:2006bx, Das:2007gm}.
The uniform-disordered phase transitions are reported on this critical line.
Our results shows that the slope of the critical line diverges in large-$N$ limit, while \cite{Panero:2006bx} reports weak $N$-dependence of the  critical line. This is because our slope of the critical line is logarithmically divergent, 
and thus it is hard to see the vertical of the critical line if $N$ is not so large.       

It is worth noticing that \eqref{rmRGE} and \eqref{rgRGE} also imply the Gaussian fixed point 
in the NCFT, where the scaling dimensions of $m_N^2$ and $g_N$ are again 2, 
because the difference between the field theory on the fuzzy sphere and the NCFT 
is just in the $N$-dependence of $b_N$ itself as in \eqref{fsb} and \eqref{NCb}, 
namely in a quantity which should be fixed in the large-$N$ limit. 
It would be intriguing aspect of our large-$N$ RG formalism 
that we can discuss a field theory on the fuzzy sphere and on the
noncommutative plane in a unified way. 

\subsection{Nontrivial fixed points}

Next we search for nontrivial fixed points. 
If there is a nontrivial fixed point, we would have nontrivial wave function renormalization there. 
In such a case if we normalize the kinetic tern canonically as in \eqref{wfrenormalization}, 
it would give rise to nontrivial $N$-dependence of $g_N$ as in \eqref{rmmaction}. 
Here we assume general leading $N$-dependence of $c(N)$ in \eqref{rmmaction} 
as $c(N)=cN^a$ and fix the value of $a$ in such a way that there exists a fixed point 
of ${\cal O}(N^0)$. 
Near such a fixed point, if any, the perturbative expansion in terms of $g_N$ can be 
in danger due to higher order interactions.
We observe that since $\rho_N^2c(N)g_NP_N^2N$ is the loop expansion parameter,  
$\rho_N^2 c(N) g_NP_N^2 N \ll 1$ is required for the loop expansion to be a good
approximation\footnote{Here we do not take further possible suppression factors 
due to $6j$-symbols into account.}. 
With this condition the third term in the left hand side of
(\ref{rmRGE}) is subleading. 
Therefore up to ${\cal O}(g_N^2)$ the RG equations (\ref{rmRGE}) and (\ref{rgRGE}) can be recast 
into 
\begin{align}
& m_{N-1}^2=b_N^2\left(m_N^2+c(N)g_NB_1(N,m_N^2)\right), \label{rmRGE2} \\
&\frac{1}{g_{N-1}}=\frac{1}{\frac{c(N)}{c(N-1)}b_{N}^2}\left(\frac{1}{g_{N}}+c(N)\rho_N^2B_2(N,m_N^2)\right). \label{rgRGE2}
\end{align}
We find a fixed point of \eqref{rmRGE2} and \eqref{rgRGE2} analogous to the Wilson-Fisher fixed point 
\begin{align}
m^2_*=-\frac{N(N-1)}{\rho_N^2}\frac{b_N^2(\frac{c(N)}{c(N-1)}b_N^2-1)}{\frac{c(N)}{c(N-1)}b_N^4-1},
\quad
g_*=\frac{N^2(N-1)^2}{2(2N-1)}
\frac{\frac{c(N)}{c(N-1)}b_N^2-1}{c(N)\rho_N^2}\left(\frac{b_N^2-1}{\frac{c(N)}{c(N-1)}b_N^4-1}\right)^2. 
\label{ntfixpt}
\end{align}
We then confirm that the loop expansion parameter $\rho_N^2 c(N) g_N
P_N^2 N$ is small around this fixed point as a consistency check. 
Using these fixed point values, we evaluate  
\begin{eqnarray}
\rho_N^2c(N)g_*P_N^2N =\frac{N}{2(2N-1)}\left(\frac{c(N)}{c(N-1)}b_N^2-1\right).
\end{eqnarray}
For arbitrary $a$, this is of ${\cal O}(N^{-1})$ and negligible
in the large-$N$ limit.
Therefore our
fixed point \eqref{ntfixpt} is stable against higher order loop
corrections, and 
this also suggests that the fixed point will not receive
much effects in general from higher order interactions induced by the RG
transformation. 
Such feature can be understood as the locality of our
RG transformation itself.

We linearize  \eqref{rmRGE2} and \eqref{rgRGE2}
around the fixed point. 
The linear perturbations obey  
\begin{align}
&\delta m^2_{N-1}=b_N^2\left(2-\frac{c(N)}{c(N-1)}b_N^2\right)\delta
m^2_{N}
+\left( 1-b_N^2 \right) \frac{m^2_*}{g_*}\delta g_N, 
\label{rmRGE3}
\\
&\delta g_{N-1}=\frac{1}{\frac{c(N)}{c(N-1)}b_N^2}\delta
g_{N}+\left(1-\frac{1}{\frac{c(N)}{c(N-1)}b_N^2}\right)\frac{\left(1-b_N^2\right)}{b_N^2}
\frac{m^2_*}{(2N-1)c(N)}\delta m^2_N.
\label{rgRGE3}
\end{align} 
The scaling dimensions are determined from the eigenvalues
of these linear perturbation equations.
Note that off-diagonal terms in these equations are of 
${\cal O}(N^{-1})$.
Thus to the leading order,
the eigenvalues for $\delta m^2$ and $\delta g$ are
\begin{eqnarray} 
\delta m^2 \;: \;b_N^2\left(2-\frac{c(N)}{c(N-1)}b_N^2\right), \qquad 
\delta g \;:\; \left(\frac{c(N)}{c(N-1)}b_N^2\right)^{-1}.
\label{eigenvalue}
\end{eqnarray}
with ${\cal O}(N^{-2})$ corrections. 
As concrete examples, we discuss scaling dimensions of these linear perturbations in both
cases of fuzzy sphere and NCFT, separately. 
\\

\noindent
{\bf fuzzy sphere case:} 
Substituting $\rho_N^2\simeq N^2\alpha^2/4$ where $\alpha^2=\alpha_N^2$ 
is fixed in the RG, we find 
\begin{align}
m^2_*\rightarrow -\frac{4}{\alpha^2}\frac{a+2}{a+4}, \quad 
g_*\rightarrow \frac{4}{c\alpha^2}\frac{a+2}{(a+4)^2}N^{-a} \quad
(N\rightarrow\infty),
\end{align} 
where we have used \eqref{fsb}. 
We choose $a=0$ to have $g_*$ of ${\cal O}(1)$, then 
$(m^2_*, g_*)\rightarrow (-\frac{2}{\alpha^2}, \frac{1}{2c\alpha^2})$. 
Since $b_N\simeq 1+1/N$ from \eqref{fsb}, the scaling dimensions of the linear perturbations are obtained from
coefficients of $1/N$ terms of the eigenvalues (\ref{eigenvalue}).
They are $0$ for $\delta m^2$ and $-2$ for $\delta g$.
Namely there are one marginal and one irrelevant perturbations
to this order in this large-$N$ limit.  
\\ 

\noindent
{\bf NCFT case:} $\rho_N^2=N\theta/2$ from \eqref{thetadef} yields 
\begin{align}
m^2_*\rightarrow -\frac{2N}{\theta}\frac{a+1}{a+2}, \quad
g_*\rightarrow \frac{N^{1-a}}{2c\theta}\frac{a+1}{(a+2)^2} \qquad
(N\rightarrow\infty), 
\end{align}
where we have utilized \eqref{NCb}. 
We set $a=1$ to have $g_*$ of ${\cal O}(1)$, 
 then $(m^2_*, g_*)\rightarrow (-\frac{4N}{3\theta}, \frac{1}{9c\theta})$. 
Although the position of $m^2_*$ goes to negative infinity as $N$ goes
large,
the condition for higher loop corrections to be suppressed
is still satisfied.
Note that now the rescaling factor is given as $b_N^2=1+\frac{1}{N}$. 
Thus the scaling dimensions of linear perturbations can be read off as
$-2$ for $\delta m^2$ and $-4$ for $\delta g$. 
This IR fixed point is stable in this large-$N$ limit.\\

\noindent
{\bf large $\rho_N^2$ limit:} When  $\rho_N^2m^2_N \gg N^2$,
the factor from the propagator $P_N$ in
$B_1(N,m_N^2)$ and $B_2(N,m_N^2)$ can be approximated
by $\rho_N^{-2}m_N^{-2}$.
Then we easily observe from \eqref{rgRGE2}
how the position of the
fixed point depends on $N$ as (for simplicity, we choose $c(N)=1$)
\begin{eqnarray}
m^4_*=\frac{2(2N-1)}{b_N^2-1}\frac{g_*}{\rho_N^2}.
\label{fixedlinelr}
\end{eqnarray}
This suggests a phase transition which occurs on such points. 
Indeed, the so-called disordered--matrix phase transition is observed on such  points in numerical simulations
\cite{Martin:2004un, Panero:2006bx, Das:2007gm}.
In \cite{Steinacker:2005wj} this phase transition is explained by
a topology change of matrix eigenvalue distribution. 
In these previous works, the matrix model action is given as
\begin{eqnarray}
S=\frac{4\pi}{N}\tr(\phi[L_i, [L_i, \phi]]+rR^2\phi^2+\lambda R^2\phi^4).
\end{eqnarray}
The relation between $(R,r,\lambda)$ and our $(\rho_N, m^2_N, g_N)$ is
determined as  $\rho_N=R$, $m^2_N=r$, $g_N=\frac{\lambda}{4\pi}$.
In \cite{Steinacker:2005wj}, noncommutativity is fixed as  $R^2=\frac{N\theta}{2}$, thus we adopt 
$b_N^2=\frac{N}{N-1}$ as in \eqref{NCb} for a comparison.  
Then the equation \eqref{fixedlinelr} becomes
\begin{eqnarray}
\frac{r}{N}=\pm \frac{\sqrt{\lambda}}{\sqrt{\pi}R}=\pm 0.564\frac{\sqrt{\lambda}}{R} \quad (N\rightarrow \infty).
\label{fixedlinelr2}
\end{eqnarray}
On the other hand, the position of the critical point on $\lambda=1$
line is calculated numerically in \cite{Martin:2004un}, 
and also estimated by an analytical method in
 \cite{Steinacker:2005wj}. 
Their results are 
\begin{align}
&\mbox{observed in \cite{Martin:2004un}}: \quad  \frac{r}{N}=-\frac{0.56}{R}, \nn \\
&\mbox{estimated in  \cite{Steinacker:2005wj}}: \quad 
\frac{r}{N}=\pm \frac{3}{2\sqrt{\pi}}\frac{1}{R}=\pm\frac{0.846}{R}.  \nn
\end{align}      
Our result \eqref{fixedlinelr2} have a good agreement with the numerical simulation\footnote{If we adopt 
$b_N^2=\frac{N}{N-2}$ as in \eqref{fsb}, the number in front of $\frac{1}{R}$ in (\ref{fixedlinelr2}) differs by $2^{-1/2}$.}.
It strongly suggests that our formula \eqref{fixedlinelr} describes the disordered--matrix phase transition points. Analysis of critical behavior of various order parameters will provide further evidence.
And the phase structures of this matrix model are
further investigated numerically in \cite{Das:2007gm}.
It would be interesting to understand these structures by our RG methods.

\subsection{Two-dimensional gravity}
\label{subsec:2dqg}
As a final application of our large-$N$ RG, in this section let us consider the two-dimensional gravity. 
For comparison we start from the same action as in the original large-$N$ RG in \cite{Brezin:1992yc}
and apply ours, namely, perform the large-$N$ RG by a different basis of $N\times N$ matrices 
from that in \cite{Brezin:1992yc}. 
Then we compare our results with those in \cite{Brezin:1992yc}. 

We begin with the one-matrix model \eqref{onemm} defining the two-dimensional quantum gravity 
\begin{align}
S_N=N\trn{N}{\frac12\phi_N^2+\frac{g}{4}\phi_N^4}, 
\label{onemm2}
\end{align}
here dropping the kinetic term $\frac12\Lap{\phi_N}^2$ would be essential for the two-dimensional gravity, 
because by doing so we have the $U(N)$ gauge symmetry in \eqref{onemm2} which would be 
somehow related to the diffeomorphism invariance.
Then it is straightforward to repeat our formalism. Two important differences from before 
are that now we fix the wave function renormalization as 
\begin{align}
N\trn{N}{\phin^2}=(N-1)\trn{N-1}{\wphi^2},
\end{align}
instead of \eqref{wfrenormalization}, and that the propagator becomes 
\begin{align}
P_N=\frac{1}{N^2}. 
\end{align}
With these in mind, it is easy to derive the RG equation for $g_N$ again as 
\begin{align}
g_{N-1}=g_N\left(1-\frac2N\left(1+6g_N\right)\right)+{\cal O}(g_N^2),
\end{align}
from which we obtain as in \eqref{BZbeta} 
\begin{align}
\beta(g)=-2g-12g^2.
\end{align}
Therefore we again find a nontrivial fixed point $g_*=-1/6$ and a critical exponent there 
$\gamma_1=2/\beta'(g_*)=1$. Although $\gamma_1$ is worse than that obtained in \cite{Brezin:1992yc}, 
they are still comparable with the exact results $g_*=-1/12$ and $\gamma_1=5/2$. 
There may be several reasons why our RG becomes worse than in \cite{Brezin:1992yc} for the two-dimensional 
gravity. First we notice that our formalism respects the rotational symmetry $SO(3)$. It is evident that 
this property fits in field theory on the (fuzzy) sphere, but not in the two-dimensional gravity describing 
random surfaces. Another reason could be that in our formalism we have dropped contributions from the antipode matrix, 
while in \cite{Brezin:1992yc} all ${\cal O}(g)$ contributions in the systematic
$1/N$ expansion are taken into account.

\section{Conclusions and discussions}
\label{sec:discussions}
In this paper we propose a new large-$N$ RG based on the fuzzy sphere. 
It has a nice analogy with the usual RG 
in some aspects
such as the locality of the RG transformation and the derivative expansion. 
It also reveals the interesting features in matrix models as the
appearance of multi-trace operators, the antipode transformation, and
renormalization of noncommutativity. 
Fixed point analysis gives 
consistent results for the Gaussian fixed point of field theory on the fuzzy sphere, or the NCFT 
on the two-dimensional plane. 
A nontrivial fixed point is also found and their properties are discussed. 
There is a critical line if the radius of the sphere is large.
Our RG equation provides an expression for the critical line which well agrees with the numerical simulation.      
A comparable value of the critical exponent in the two-dimensional gravity 
can be also obtained. 

Since it seems straightforward to apply our formalism to multi-matrix models, it is desirable to apply 
it to the large-$N$ limit of more interesting models, related to string theory, like the Chern-Simons type 
matrix models as in e.g. \cite{Iso:2001mg, Azuma:2005bj, Ishii:2008ib}. 
Our formalism is expected to reveal 
some aspects of them, in particular, on their universal properties. 
It is also anticipated that our large-$N$ RG 
also sheds light on the large-$N$ limit 
of supersymmetric gauge theories on higher dimensional sphere 
that have attracted much attention recently. 
It would be likely that in order to clarify vast universality class of large-$N$ gauge theories 
suggested by the large-$N$ reduced models \cite{Eguchi:1982nm}, 
we would need formalism based on the RG like ours. 
We mention that it would be also interesting to examine a relation between our large-$N$ RG  
and a master field approach on the fuzzy sphere \cite{Kuroki:1998rx}.
\\

Before closing the paper, we discuss two issues which would be
considerable for further understanding of the matrix RG method in
the present paper.

\subsubsection*{Fuzzy torus}     
One may think that our large-$N$ RG is also available on the fuzzy torus, where 
the algebra of functions is more tractable. In fact, an $N\times N$ matrix $\phi_N$ can be 
expanded analogously to the Fourier expansion as (assuming odd $N$)
\begin{align}
\phi_N=\frac{1}{(2\pi)^2}\sum_{n_1,n_2=-\frac{N-1}{2}}^{\frac{N-1}{2}}\phi_{n_1\,n_2}\omega^{\frac12n_1n_2}U^{n_1}V^{n_2},
\end{align} 
where $U$ and $V$ are the standard 't Hooft matrices with rank $N$ satisfying $UV=\omega^{-1}VU$ with 
$\omega=\exp(2\pi i/N)$. Then each vertex of a field theory on the fuzzy torus has a factor 
imposing the momentum conservation modulo $N$: $\delta^{(N)}(\sum_rn_1^{(r)})\delta^{(N)}(\sum_rn_2^{(r)})$. 
In our large-$N$ RG, it is natural to integrate first over $(4N-4)$ modes with the maximal momentum 
of $n_1$ or $n_2$, namely, over $\phi_{n_1\,n_2}$ with $|n_1|$ or
$|n_2|=(N-1)/2$, and then to rewrite the resulting action as an
$(N-2)\times (N-2)$ matrix model.
However, in the $(N-2)\times(N-2) $ matrix model the momentum
conservation should hold in each vertex modulo $N-2$. 
For example, if a vertex consists only of in-modes with
$|n_1|,|n_2| < (N-1)/2$, it remains intact in the RG,
and hence there the momentum is conserved modulo $N$, not $N-2$.  
This serious problem would be bypassed by considering the large-$N$ RG
not by $N\rightarrow N-2$,
but by $N\rightarrow N/2$ as the block spin transformation. We will
report our study in this direction elsewhere.

\subsubsection*{Antipode, UV/IR mixing, and noncommutative anomaly}

Another interesting aspect in our formalism is the appearance of the
antipode transformation. Here we point out an interesting connection
between the antipode matrix and the nonplanar diagram. If we draw
nonplanar diagrams by the standard double line notation, we find
that they resemble annulus diagrams appearing at one-loop of open
string theory. 
They have two edges from which external lines
emanate. Our intriguing observation up to ${\cal O}(g_N^2)$ is that 
if we identify matrices brought by external lines from one edge as ``usual'' ones, 
external lines from another edge provide the antipode matrices. 
Namely, there is a connection between the Feynman diagram and the positions on
the (fuzzy) sphere. We hope that further analysis of nonplanar
diagrams makes clear the origin of the antipode transformation.

Related to this, in \cite{Vaidya:2001bt} it is claimed that such
antipode degrees of freedom trigger the UV/IR mixing
\cite{Minwalla:1999px}. After \cite{Vaidya:2001bt}, it is rigorously
argued in \cite{Chu:2001xi} that there is no UV/IR mixing on the fuzzy
sphere for finite $N$ based on the one-loop effective action including a nonplanar
diagram, and the IR singularity seen in field theory on the noncommutative
plane is reproduced as the large-$N$ limit of  a term arising from a nonplanar diagram, 
that the authors call noncommutative anomaly (it is regular for finite $N$). 
As seen below in detail, the appearance of the antipode matrix reported in this
paper can be understood as a part of such a noncommutative anomaly for
finite $N$.

Here we make a detailed comparison of $\vev{ V_2^{NP}}$ between that
in this paper and in \cite{Chu:2001xi}.
The authors of \cite{Chu:2001xi} use the following formula in calculating $\vev{V_2^{NP}}$ 
\begin{eqnarray}
\label{Chu_formula}
\left\{ \begin{array}{ccc}
L&L& l\\
L&L& J
\end{array}\right\}
\simeq\frac{(-1)^{2L+l+J}}{2L}P_{l}
\left(
1-\frac{J^2}{2L^2} \right) 
\,.
\end{eqnarray}
The noncommutative anomaly for finite $N$
is obtained as the difference between $\vev{V_2^{NP}}$ and $\vev{V_2^P}$ 
with the internal mode integrated over the whole region $0 \leq J
\leq 2L$ by use of this formula.
On the other hand, our antipode matrix originates in
$\vev{V_2^{NP}}_0$ with only the $J=2L$ mode integrated
and with \eqref{6jexp1} applied.
The term with $J=2L$ in \eqref{Chu_formula} indeed
reproduces the first term\footnote{%
Note that \eqref{Chu_formula} and \eqref{6jexp1} are 
different approximation formulas even when $J=2L$,
and
the ``derivative corrections'' in  \eqref{6jexp1} are not visible
by use of \eqref{Chu_formula}.}
 in the formula \eqref{6jexp1}. 
Therefore, the antipode effect of $\vev{V_2^{NP}}_0$ in this paper can
be regarded as the maximum momentum part of the noncommutative anomaly
discussed in \cite{Chu:2001xi}.
It is then preferable that the other appearances of the antipode matrix
can also be interpreted in a similar way.
If we can repeatedly integrate out the highest modes, the total antipode
effect will eventually, in principle, reproduce the noncommutative
anomaly for finite $N$. Our RG approach would be helpful toward such a
problem, as discussed next.

Since as shown in \cite{Chu:2001xi} the noncommutative anomaly
develops the IR singularity in the large-$N$ NCFT limit \eqref{NCFTlimit} 
and triggers the UV/IR mixing, the above observation implies that around a fixed point corresponding to the large-$N$ 
limit in our RG, interactions involving the antipode matrix would be relevant and that analysis of them there 
would reveal universal nature of the noncommutative anomaly and the
UV/IR mixing.
More generally, an important question from our study is on existence of nontrivial
theory with the antipode matrix. 
According to the spirit of the RG, we are led to consider a new matrix model 
like
\begin{align}
S_N=N\trn{N}{-\frac12[L_i,\phi]^2-\alpha[L_i,\phi][L_i,\phi^A]+\frac{\rho_N^2m^2}{2}\phi^2+\rho_N^2\widetilde m^2\phi\phi^A+\cdots},
\end{align} 
and look for a nontrivial fixed point of our large-$N$ RG. 
For example, the fixed point analysis of $\widetilde{m}^2$ is expected to be useful 
for understanding the UV/IR mixing. \\

Finally, we comment on the observations made in
\cite{Minwalla:1999px, VanRaamsdonk:2000rr}. In these papers
the UV/IR mixing is understood as the exchanges of light degrees of
freedom which are called closed string modes. This nicely fits with stringy
interpretation of a nonplanar diagram. Recall that the 1-loop correction 
of the scalar field 2-point function is UV divergent in the 2-dimensional continuum theory 
and that if we try to regularize it via the noncommutativity as in the fuzzy sphere or in the noncommutative plane, 
a nonplanar diagram in general yields the noncommutative anomaly, or the UV/IR mixing, in the large-$N$ limit.  
Then the IR singularity there can be regarded as the effect of the propagation of the closed sting mode. 
Therefore it is natural to expect that even for finite $N$ a nonplanar diagram of the 2-point function 
in the matrix model on fuzzy sphere has a counterpart of such a closed string mode. 
In fact, the formula \eqref{V2NPtemp3} can be written as the exchange of particles with propagator
\begin{eqnarray}
\Delta(p)=\int d\mu^2\;\lambda(\mu^2)\frac{1}{p\circ p/\rho^2+\mu^2}, \quad 
\lambda(\mu^2)=\delta \left( \mu^2-\frac{1}{N^2} \right),
\label{Delta}
\end{eqnarray} 
following the expression in \cite{VanRaamsdonk:2000rr}.
Here identification is made as
\begin{eqnarray}
l(l+1)=p^2\rho^2, \quad  
\frac{\rho^4}{N^3}p^2= p\circ p,
\end{eqnarray}
and $\rho$ is the radius of sphere. 
The spectral function $\lambda(\mu^2)$ in  \cite{VanRaamsdonk:2000rr}
has a uniform distribution which vanishes below $1/(\rho\Lambda)^2$, where $\Lambda$ is a UV cutoff of Wilsonian RG in NCFT 
and $\rho$ is replaced by a parameter with dimensions of squared length in the reference. 
On the other hand, the spectral function in (\ref{Delta}) is the delta function. This is
because we calculated a one-step RG flow $N \rightarrow N-1$. 
It is shown that matrix models on the fuzzy sphere are free from the IR divergence 
for finite $N$ \cite{Chu:2001xi}. 
In fact, this feature can be also seen in (\ref{Delta}). The spectral function does not contain
massless modes unless $N$ is infinity. An IR regularization of the closed string picture is
realized in this way. \\

\section*{Acknowledgments}
We are also grateful to H. Kawai, S. Iso, A. Miwa, H. Shimada, and F. Sugino for fruitful discussions. 
The work of T.K. was supported in part by Rikkyo University Special
Fund for Research. 
The work of S.K. was supported by NSC100--2811--M--029--003.
D.T. was supported by NSC 100-2811-M-029-002.   

\appendix 

\section{Useful formulas of $3nj$-symbols}
\label{app:3nj}
In this appendix we enumerate useful formulas for our study from \cite{Var}. 
The Latin and corresponding Greek indices denote angular momenta 
and magnetic quantum numbers (projection of angular momentum), respectively. 
For example, the magnetic quantum number $\alpha$ runs 
$-a\leq \alpha \leq a$ for the angular momentum $a$. 
In the following we define 
\begin{align}
\{abc\}=
\left\{
\begin{array}{ccl}
1 & & \quad |a-b|\leq c\leq a+b \\
0 & & \quad \text{otherwise}
\label{triangular}
\end{array}
\right., 
\end{align}
namely, $\{abc\}$ does not vanish only if $a$, $b$, and $c$ 
satisfy the triangular conditions which are symmetric 
under the interchange of variables. 
\subsection{Identities}
\begin{align}
\sum_{\psi\kappa}(-1)^{p-\psi+q-\kappa}
\bepr
a & p & q \\
-\alpha & \psi & \kappa
\eep
\bepr
p & q & a' \\
-\psi & -\kappa & \alpha'
\eep
=
\frac{(-1)^{a+\alpha}}{2a+1}
\{apq\}
\delta_{aa'}\delta_{\alpha\alpha'}.
\label{p453(3)}
\end{align}
\begin{align}
\sum_{q\kappa}(-1)^{q-\kappa}(2q+1)
\bepr
a & b & q \\
-\alpha & -\beta & \kappa
\eep
\left(
\begin{array}{rll}
q & a & b \\
-\kappa & \alpha' & \beta'
\end{array}
\right)
=(-1)^{a+\alpha+b+\beta}
\delta_{\alpha\alpha'}\delta_{\beta\beta'}.
\label{p453(4)}
\end{align}
\begin{align}
\sum_{\kappa\psi\rho}
(-1)^{p-\psi+q-\kappa+r-\rho}
\bepr
p & a & q \\
\psi & \alpha & -\kappa
\eep
\bepr
q & b & r \\
\kappa & \beta & -\rho
\eep
\bepr
r & c & p \\
\rho & \gamma & -\psi
\eep
=
\bepr
a & b & c \\
-\alpha & -\beta & -\gamma
\eep
\beBr
a & b & c \\
r & p & q
\eeB.
\label{p454(6)}
\end{align}
\begin{align}
&\sum_{\psi\kappa\rho\sigma}
(-1)^{p-\psi+q-\kappa+r-\rho+s-\sigma}
\bepr
p & a & q \\
\psi & \alpha & -\kappa
\eep
\bepr
q & b & r \\
\kappa & \beta & -\rho
\eep
\bepr
r & c & s \\
\rho & \gamma & -\sigma
\eep
\bepr
s & d & p \\
\sigma & \delta & -\psi
\eep \nn \\
&=
(-1)^{s-a-d-q}
\sum_{x\xi}(-1)^{x-\xi}(2x+1)
\bepr
a & x & d \\
\alpha & -\xi & \delta
\eep
\bepr
b & x & c \\
\beta & \xi & \gamma 
\eep
\beBr
a & x & d \\
s & p & q
\eeB
\beBr
b & x & c \\
s & r & q
\eeB
\label{p455(10)}
\\
&=
(-1)^{2r-s-d-a-q}
\sum_{x\xi}(-1)^{x-\xi}(2x+1)
\bepr
a & x & c \\
\alpha & -\xi & \gamma
\eep
\bepr
b & x & d \\
\beta & \xi & \delta
\eep
\beBr
a & p & q \\
x & d & b \\
c & s & r
\eeB.
\label{p455(11)}
\end{align}
\begin{align}
\sum_X(-1)^{p+q+X}(2X+1)
\beB
a &  b & X \\
c & d & p
\eeB
\beB
c & d & X \\
b & a & q
\eeB
=
\beB
c & a & q \\
d & b & p
\eeB.
\label{p463(8)}
\end{align}
\begin{align}
\sum_X(-1)^{2X}(2X+1)
\begin{Bmatrix}
a & b & X\\
c & d & p
\end{Bmatrix}
\begin{Bmatrix}
c & d & X\\
e & f & q
\end{Bmatrix}
\begin{Bmatrix}
e & f & X\\
a & b & r
\end{Bmatrix}
=
\begin{Bmatrix}
a & f & r \\
d & q & e \\
p & c & b
\end{Bmatrix}.
\label{p466(19)}
\end{align}
\subsection{Asymptotic formulas of $6j$-symbol}
If $a, b, c\gg f$ and $f$ is an arbitrary integer, 
\begin{align}
\beB
a & b & c \\
b & a & f
\eeB
\simeq
\frac{(-1)^{a+b+c+f}}{\sqrt{(2a+1)(2b+1)}}P_f(\cos\theta), 
\label{p308(16)}
\end{align}
with 
\begin{align}
\cos\theta=\frac{a(a+1)+b(b+1)-c(c+1)}{2\sqrt{a(a+1)b(b+1)}}.
\label{p308(15)}
\end{align}
If $R\gg 1$ and $a$, $b$, $c$ are arbitrary, 
\begin{align}
(-1)^{2R}
\beB
a & b & c \\
d+R & e+R & f+R
\eeB
\simeq
\frac{(-1)^{c+d+e}}{\sqrt{2R(2c+1)}}C_{a(f-e)\,b(d-f)}^{c(d-e)}, 
\label{p306(4)}
\end{align}
where $C_{a\alpha\,b\beta}^{c\gamma}$ is the Clebsch-Gordan coefficient\footnote
{We have supplied the formula presented in \cite{Var} with a necessary phase  
which was also added in \cite{Chu:2001xi}.}.\\

To investigate the asymptotic behavior of the $6j$-symbols,
it is sometimes useful to use the explicit expression due to Racah,
\begin{align}
    \begin{Bmatrix}
        A & B & C \\ a & b & c
    \end{Bmatrix}
=&
\sqrt{\Delta(A,B,C) \Delta(A,b,c) \Delta(a,B,c) \Delta(a,b,C)}
\sum_t \frac{(-1)^t (t+1)!}{f(t)}
\,,
\label{Racah_formula}
\\
\nn
f(t)=& (t-n_1)! (t-n_2)! (t-n_3)! (t-n_4)! 
(m_1 - t)!  (m_2 - t)!  (m_3 - t)! 
\,,
\\
\nn
\Delta(x,y,z) \equiv &
\frac{(x+y-z)! (y+z-x)! (z+x-y)!}{(x+y+z+1)!} \,,
\\
\nn
n_1 =& A+B+C \,,
\qquad
n_2 = A+b+c
\,, \qquad
n_3 = a+B+c \,,
\qquad
n_4 = a+b+C \,,
\\
\nn
m_1 =& A+a+B+b \,,
\qquad
m_2 = B+b+C+c \,,
\qquad
m_3 = C+c+A+a \,,
\end{align}
where the sum of $t$ runs in the region where no argument of the factorials
takes negative values, namely,
\begin{align}
  \text{max} \left( n_1 ,n_2, n_3, n_4 \right)
\leq t \leq
  \text{min} \left( m_1 ,m_2, m_3 \right)
\,.\nn
\end{align}
Several asymptotic formulas are obtained by evaluating
the factorials by the use of the Stirling's formula.

\section{Useful formulas of the fuzzy spherical harmonics}
\label{app:tlm}
In this appendix useful formulas of the $N\times N$ matrix $\tlm$ 
defined in \eqref{Tdef} are presented. We set $L=(N-1)/2$ as in the main text. 
\eqref{p454(6)}, \eqref{p455(10)} and \eqref{p455(11)} 
give formulas of the trace 
of $\tlm$'s 
\begin{align}
&\trn{N}{\tn{1}\tn{2}\tn{3}}
=N^{\frac32}\prod_{i=1}^3(2l_i+1)^{\frac12}(-1)^{2L+\sum_{i=1}^3l_i}
\bep
l_1 & l_2 & l_3 \\
m_1 & m_2 & m_3
\eep
\beB
l_1 & l_2 & l_3 \\
L & L & L
\eeB,
\label{3T} \nn \\
&\trn{N}{\tn{1}\tn{2}\tn{3}\tn{4}} \\
&=N^2\prod_{i=1}^4(2l_i+1)^{\frac{1}{2}}\sum_{lm}(-1)^{-m}(2l+1) 
\bep
l_1 & l_4 & l \\
m_1 & m_4 & m
\eep
\bep
l & l_3 & l_2 \\
-m & m_3 & m_2
\eep
\beB
l_1 & l_4 & l \\
L & L & L
\eeB
\beB
l & l_3 & l_2 \\
L & L & L
\eeB 
\label{4T1}
\\
&=N^2\prod_{i=1}^4(2l_i+1)^{\frac{1}{2}}(-1)^{l_2+l_3}\sum_{lm}(-1)^{l-m}(2l+1) 
\bep
l_1 & l_3 & l \\
m_1 & m_3 & m
\eep
\bep
l & l_2 & l_4 \\
-m & m_2 & m_4 
\eep
\beB
l_1 & l_3 & l \\
L & L & l_2 \\
L & L & l_4
\eeB. 
\label{4T2}
\end{align}
Notice that \eqref{4T1} can be also derived easily from \eqref{3T} and \eqref{Tcomp},  
and that $l_i$ ($i=1\sim 4$) on the right-hand side of \eqref{4T1} can
be cyclically interchanged, which 
reflects the cyclic symmetry of the trace. 
{}From \eqref{Tortho} and \eqref{3T} we can read off 
a fusion coefficient ${F_{l_1m_1\,l_2m_2}}^{l_3m_3}$ defined by 
\begin{align}
\tn{1}\tn{2}=\sum_{l_3m_3}{F_{l_1m_1\,l_2m_2}}^{l_3m_3}\tn{3},
\label{fuzzyfusion}
\end{align}
as 
\begin{align}
{F_{l_1m_1\,l_2m_2}}^{l_3m_3}=N^{\frac12}\prod_{i=1}^3(2l_i+1)^{\frac12}
(-1)^{2L+\sum_{i=1}^3l_i+m_3}
\bep
l_1 & l_2 & l_3 \\
m_1 & m_2 & -m_3
\eep
\beB
l_1 & l_2 & l_3 \\
L & L & L
\eeB.
\label{fusion}
\end{align}
As an application, by using the asymptotic form of the $6j$-symbol 
for $L\gg 1$ in \eqref{p306(4)} 
\begin{align}
(-1)^{2L}
\beB
a & b & c \\
L & L & L
\eeB
\simeq 
\frac{(-1)^c}{\sqrt{2L(2c+1)}}C_{a0\,b0}^{c0},
\end{align} 
and 
\begin{align}
\bep
l_1 & l_2 & l_3 \\
m_1 & m_2 & m_3
\eep
=(-1)^{l_3+m_3+2l_1}\frac{1}{\sqrt{2l_3+1}}C_{l_1\,-m_1\,l_2\,-m_2}^{l_3
  m_3},
\end{align}
\eqref{fuzzyfusion} and \eqref{fusion} reproduces the fusion of 
the usual spherical harmonics \cite{Var}
\begin{align}
Y_{l_1m_1}Y_{l_2m_2}=\sum_{l_3m_3}\sqrt{\frac{(2l_1+1)(2l_2+1)}{2l+1}}
C_{l_10\,l_20}^{l_30}C_{l_1m_1\,l_2m_2}^{l_3m_3}Y_{l_3m_3}.
\end{align}
Hence the structure constant of the fuzzy spherical harmonics 
\begin{align}
&\left[\tn{1},\tn{2}\right]=\sum_{l_3m_3}f_{l_1m_1\,l_2m_2\,l_3m_3}\tn{3}^{\dagger}, \nn \\
&f_{l_1m_1\,l_2m_2\,l_3m_3}=(-1)^{m_3}\left({F_{l_1m_1\,l_2m_2}}^{l_3\,-m_3}-{F_{l_2m_2\,l_1m_1}}^{l_3\,-m_3}\right),
\label{fuzzystrconst}
\end{align}
also coincides with 
that of the usual spherical harmonics in the large-$N$ limit. 

Furthermore, by using the fusion \eqref{fuzzyfusion} and \eqref{fusion} repeatedly,
we find 
\begin{align}
\ton{}\tn{1}\tom{}
=&N(4L+1)\sum_{{l'm'}\atop{l^{\pp}m^{\pp}}}
(2l'+1)\sqrt{(2l_1+1)(2l^{\pp}+1)}(-1)^{l_1+l^{\pp}+m'+m^{\pp}} \nn \\
&\times 
\bep
l_1 & l' & 2L \\
m_1 & -m' & m
\eep
\bep
l' & 2L & l^{\pp} \\
m' & -m & -m^{\pp}
\eep
\beB
2L & l_1 & l' \\
L & L & L
\eeB
\beB
l' & 2L & l^{\pp} \\
L & L & L
\eeB
T_{l^{\pp}m^{\pp}}.
\end{align}
Taking the sum $\sum_{m=-2L}^{2L}(-1)^m$ of both sides of this equation 
and applying \eqref{p453(3)}, we have 
\begin{align}
\sum_{m=-2L}^{2L}&(-1)^m\ton{}\tn{1}\tom{} \nn \\
=&(-1)^{2L+l_1}N(2N-1)\tn{1} 
\sum_{l'}(-1)^{l'}(2l'+1)
\beB
2L & l_1 & l' \\
L & L & L
\eeB^2,
\end{align}
where the summation over $l'$ can be also taken by utilizing \eqref{p463(8)} to yield  
\begin{align}
\sum_{m=-2L}^{2L}(-1)^m\ton{}\tn{1}\tom{}
=N(2N-1)
\beB
L & L & l_1 \\
L & L & 2L
\eeB
(-1)^{l_1}\tn{1}.
\label{similarity}
\end{align}
This equation tells us an interesting fact that making the similarity transformation 
with respect to $\ton{}$ and taking the summation over $m$ is essentially 
equivalent to performing the ``antipode" transformation $\tn{}\rightarrow (-1)^l\tn{}$, 
namely transporting a field to the antipode on the sphere.  
As a corollary, we find that  
\begin{align}
\sum_{m=-2L}^{2L}(-1)^m\ton{}\tom{}
=(2N-1)\bm 1_{N}, 
\label{coro}
\end{align}
where we have used $T_{00}=\bm 1_N$ and 
\begin{align}
\beB
L & L & 0\\
L & L & 2L
\eeB
=\frac1N.
\end{align}  
\eqref{similarity} and \eqref{coro} are of great help 
in calculating our RG.

\section{Calculations of ${\cal O}(g_N^2)$}
\label{app:Og2}
In this appendix we evaluate contributions of ${\cal O}(g_N^2)$ in the large-$N$ RG 
shown in the second line in \eqref{RGexp}. 
First we consider $\vev{V_2^2}_c$ that contributes to vertex correction 
and then turn to the other terms giving rise to mass correction.  
\subsection{Vertex correction}
$V_2=V_2^P+V_2^{NP}$ as in \eqref{vertices} and the only possible planar diagram in $\vev{V_2^2}_c$ 
arises in $\vev{{V_2^P}^2}_c$ calculated in \eqref{V2P2ptemp}. Furthermore, 
in the low energy regime where the momentum $l$ of $\pin{}$ is much smaller than the cutoff $2L$,  
we can perform ``derivative"  expansion of it as in \eqref{V2P2pexp} to get \eqref{V2P2p}. 
Similarly we can evaluate the nonplanar diagram in $\vev{{V_2^P}^2}_c$ given as the second term 
in \eqref{V2P2temp} and obtain a similar result to \eqref{V2P2ptemp} 
\begin{align}
\left.\vev{{V_2^P}^2}_c\right|_{\text{nonplanar}}=&P_N^2(2N-1)^2N
\sum_{lm}
\beB
2L & 2L & l \\
L & L & L
\eeB^2 
(-1)^l\trn{N}{\phin^2\tlm}\trn{N}{\phin^2\tlm^\dagger{}}.
\label{V2P2nptemp}
\end{align}
Here we note that 
\begin{align}
(-1)^l\trn{N}{\phin^2\tlm}=\trn{N}{{\phinA}^2\tlm},
\end{align}
which can be readily shown directly from the expression of the trace of three $\tlm$'s 
given in \eqref{3T} by using the well-known property of the $3j$-symbol, 
or from the proposition proved in the appendix \ref{app:antipode}. 
Then in the low energy regime the same derivation as in \eqref{6jexp}$\sim$\eqref{V2P2p}
leads 
\begin{align}
&\left.\vev{{V_2^P}^2}_c\right|_{\text{nonplanar}} \nn \\
&=P_N^2(2N-1)N
\left[\trn{N}{\phin^2{\phinA}^2}-\frac{1}{2N}\trn{N}{\Lap{\phin^2}{\phinA}^2}+{\cal O}\left(\frac{1}{N^2}\right)\right].
\end{align}
On the other hand, from \eqref{vertices} we obtain 
\begin{align}
&\vev{V_2^PV_2^{NP}}_c \nn \\
&=\frac12\sum_{m_1,m_2,m'_1,m'_2}
\left(\vev{\pon{1}\pond{2}}_0\vev{\pon{2}\pond{1}}_0+\vev{\pon{1}\pond{1}}_0\vev{\pon{2}\pond{2}}_0\right) \nn \\
&\hspace{2.5cm}\times\trn{N}{\phin^2\ton{1}\ton{2}}\trn{N}{\phin\tond{1}\phin\tond{2}} \nn \\
&=P_N^2\sum_{m,m'}(-1)^{m+m'}\trn{N}{\phin^2 T_{2Lm} T_{2km'}}
\trn{N}{\phin T_{2L-m'} \phin T_{2L-m}},
\end{align}
where the two kinds of the contractions yielding nonplanar diagrams 
turn out to be equivalent from the cyclicity of the trace. By expanding $\phin^2$ as 
$\phin^2=\sum_{lm}\phi^{\text{in}\,(2)}_{lm}\tlm$ and using the trace formula in \eqref{3T} and in \eqref{4T1}, 
we have 
\begin{align}
&\vev{V_2^PV_2^{NP}}_c \nn \\
&=P_N^2(2N-1)^2N^{\frac72}\sum_{m,m'}(-1)^{m+m'}
\sum_{{(l_r,m_r)\in\Li} \atop {r=1\sim 3}}\phi^{\text{in}\,(2)}_{l_1m_1}\pin{2}\pin{3} \nn \\
&\times (-1)^{2L+l_1} \prod_{r=1}^3(2l_r+1)^{\frac12}
\begin{pmatrix}
l_1 & 2L & 2L \\
m_1 & m & m'
\end{pmatrix}
\begin{Bmatrix}
l_1 & 2L & 2L \\
L & L & L
\end{Bmatrix} \nn \\
&\times\sum_{l^{\pp}m^{\pp}}(-1)^{m^{\pp}}(2l^{\pp}+1)
\begin{pmatrix}
l_2 & 2L & l^{\pp} \\
m_2 & -m & -m^{\pp}
\end{pmatrix}
\begin{pmatrix}
l ^{\pp} & l_3 & 2L \\
m^{\pp} & m_3 & -m'
\end{pmatrix}
\begin{Bmatrix}
l_2 & 2L & l^{\pp} \\
L & L & L
\end{Bmatrix}
\begin{Bmatrix}
l^{\pp} & l_3 & 2L \\
L & L & L
\end{Bmatrix}.
\label{V2PV2NPtemp}
\end{align}
Here we can take the summations over $m$, $m'$, and $m^{\pp}$ according to \eqref{p454(6)}, 
while the sum on $l^{\pp}$ can be also taken by \eqref{p466(19)}. We thus get 
\begin{align}
\vev{V_2^PV_2^{NP}}_c 
=&P_N^2(2N-1)^2N^{\frac72}
\sum_{{(l_r,m_r)\in\Li} \atop {r=1\sim 3}}\phi^{\text{in}\,(2)}_{l_1m_1}\pin{2}\pin{3}
\prod_{i=1}^3(2l_i+1)^{\frac12}(-1)^{l_3} \nn \\
&\times\begin{pmatrix}
l_1 & l_2 & l_3 \\
m_1 & m_2 & m_3
\end{pmatrix}
\begin{Bmatrix}
l_1 & 2L & 2L \\
L & L & L
\end{Bmatrix} 
\begin{Bmatrix}
l_2 & l_3 & l_1 \\
L & L & 2L \\
L & L & 2L
\end{Bmatrix}.
\label{V2PV2NPtemptemp}
\end{align}
We next use the cyclic symmetry of the $9j$-symbol and utilize \eqref{p466(19)} once again. 
Then the $9j$-symbol can be rewritten as 
\begin{align}
\begin{Bmatrix}
l_2 & l_3 & l_1 \\
L & L & 2L \\
L & L & 2L
\end{Bmatrix}
=&
\begin{Bmatrix}
l_1 & l_2 & l_3 \\
2L & L & L \\
2L & L & L
\end{Bmatrix} 
=\sum_x(-1)^{2x}(2x+1)
\begin{Bmatrix}
l_1 & l_2 & l_3 \\
L & L & x
\end{Bmatrix}
\begin{Bmatrix}
2L & L & L \\
l_2 & x & L
\end{Bmatrix}
\begin{Bmatrix}
2L & L & L \\
x & l_1 & 2L
\end{Bmatrix}.
\label{9jdecomp}
\end{align}
Note that in contrast to $l^{\pp}$ in \eqref{V2PV2NPtemp}, 
here the selection rules of the $6j$-symbols impose 
$L\leq x \leq L+\text{min}(l_1, l_2)$ and $x\in\bm Z$.  
Thus setting $x=L+m$, we have 
\begin{align}
(-1)^{2L} \sum_{m=0}^{\text{min}(l_1,l_2)}(N+2m)
\begin{Bmatrix}
l_1 & l_2 & l_3 \\
L & L & L+m
\end{Bmatrix}
\begin{Bmatrix}
2L & L & L \\
l_2 & L+m & L
\end{Bmatrix}
\begin{Bmatrix}
2L & L & L \\
L+m & l_1 & 2L
\end{Bmatrix}.
\label{9jdecomp2}
\end{align}
Three $6j$-symbols above can be explicitly evaluated via the Racah
formula (\ref{Racah_formula})
and it can be further approximated by the Stirling formula 
in the low energy regime $l_r\ll L$ ($r=1\sim 3$). 
For instance, 
\begin{align}
&\begin{Bmatrix}
2L & L & L \\
l & L+m & L
\end{Bmatrix} \nn \\
&=\frac{(-1)^m}{\sqrt{m!}}\sqrt{\frac{(l+m)!}{(l-m)!}}
\sqrt{\frac{((2L)!)^2(2L+m)!(2L-m)!(4L+m+1)!}{(4L+1)!(2L-l)!(2L+l+1)!(2L+m-l)!(2L+m+l+1)!}} \nn \\
&=\frac{(-1)^m}{\sqrt{m!}}\sqrt{\frac{(l+m)!}{(l-m)!}}\frac{1}{2L^{\frac{m}{2}+1}}
\left(1-\frac{8l(l+1)-m(m-5)+8}{16L}+{\cal
    O}\left(\frac{1}{L^2}\right)\right)
\,,
\end{align}
where we note that terms with higher $m$ are suppressed as $L^{-1-\frac{m}{2}}$ when $L\gg 1$. 
We can evaluate the last $6j$-symbol in \eqref{9jdecomp2} in a similar
manner,
\begin{align}
&\begin{Bmatrix}
2L & L & L \\
L+m & l & 2L
\end{Bmatrix} 
&=\frac{(-1)^l}{\sqrt{m!}}\sqrt{\frac{(l+m)!}{(l-m)!}}\frac{\sqrt{2}}{\left(4L
  \right)^{\frac{m}{2}+1}}
\left(1-\frac{2l(l+1)+m(5m+7)+6}{16L}+{\cal
    O}\left(\frac{1}{L^2}\right)\right)
\,, \nn
\end{align}
and this is also suppressed as $L^{-1-\frac{m}{2}}$.  
Therefore, in the $1/L$ expansion, it is sufficient to consider $m=0$ and $1$ in
the summation.
As for the remaining $6j$-symbol, we can relate $m=1$ one with $m=0$
case for $l_i\ll L$ ($i=1\sim 3$) 
\begin{align}
\beB
l_1 & l_2 & l_3 \\
L & L & L+1
\eeB
\simeq
\frac{l_3(l_3+1)-l_1(l_1+1)-l_2(l_2+1)}{2\sqrt{l_1(l_1+1)l_2(l_2+1)}}
\beB
l_1 & l_2 & l_3 \\
L & L & L
\eeB,
\end{align}
by use of the asymptotic formula \eqref{p306(4)} and a recursion relation for the Clebsch-Gordan coefficient
\begin{align}
C^{c0}_{a1b-1}=C^{c0}_{a0b0}\frac{c(c+1)-a(a+1)-b(b+1)}{2\sqrt{a(a+1)b(b+1)}}.
\end{align}  

Combining these, we obtain 
\begin{align}
&\sum_{m=0}^{\text{min}(l_1, l_2)}(N+2m)
\begin{Bmatrix}
l_1 & l_2 & l_3 \\
L & L & L+m
\end{Bmatrix}
\begin{Bmatrix}
2L & L & L \\
l_2 & L+m & L
\end{Bmatrix}
\begin{Bmatrix}
2L & L & L \\
L+m & l_1 & 2L
\end{Bmatrix} \nn \\
&=\frac{(-1)^{l_1}}{\sqrt{2}N}
\beB
l_1 & l_2 & l_3 \\
L & L & L
\eeB
\left[
1-\frac1N\left(
-\frac14 l_1(l_1+1)
+\frac12l_2(l_2+1)
+\frac12l_3(l_3+1)
-\frac14\right)
+{\cal O}\left(\frac{1}{N^2}\right)
\right] \,.
\label{msum}
\end{align}
Plugging \eqref{msum} into \eqref{V2PV2NPtemptemp}, 
together with (\ref{6jexp}),
we see that 
the remaining factors are exactly of the form of \eqref{3T} and finally obtain 
\begin{align}
&\vev{V_2^PV_2^{NP}}_c \nn \\
&=P_N^2(2N-1)N
\biggl[\trn{N}{\phin^3\phinA}
-\frac{1}{2N}\trn{N}{\Lap{\phin}\phin^2\phinA} \nn \\
&\hspace{5.9cm}-\frac{1}{2N}\trn{N}{\Lap{\phinA}\phin^3}+{\cal O}\left(\frac{1}{N^2}\right)\biggr].
\end{align}
The calculation of $\vev{{V_2^{NP}}^2}_c$ is similar. The two ways of contractions again give the same 
result as 
\begin{align}
\vev{{V_2^{NP}}^2}_c
=& 
\frac{P_N^2}{2} \sum_{m_3,m_4}(-1)^{m_3+m_4}
\trn{N}{\phin\ton{3}\phin\ton{4}}\trn{N}{\phin\tom{4}\phin\tom{3}} \nn \\
=&\frac{P_N^2}{2}(2N-1)^2 N^4
\sum_{{(l_r,m_r), (l'_r,m'_r)\in\Li}\atop{r=1,2}}
\prod_{r=1}^2\left((2l_r+1)^{\frac12}\pin{r}(2l'_r+1)^{\frac12}\pind{r}\right)(-1)^{l_2+l'_2} \nn \\
&\times 
\sum_{lm}(-1)^{-m}(2l+1)
\begin{pmatrix}
l_1 & l_2 & l \\
m_1 & m_2 & -m
\end{pmatrix}
\begin{pmatrix}
l'_1 & l'_2 & l \\
m'_1 & m'_2 & m
\end{pmatrix}
\begin{Bmatrix}
l_1 & l_2 & l \\
L & L & 2L \\
L & L & 2L
\end{Bmatrix}
 \begin{Bmatrix}
l'_1 & l'_2 & l \\
L & L & 2L \\
L & L & 2L
\end{Bmatrix}, 
\label{V2NP2temp}
\end{align}
where we have used the formula for the trace of four $\tlm$'s 
in terms of $9j$-symbols \eqref{4T2} instead of $6j$-symbols \eqref{4T1}. 
Then we apply a similar evaluation from \eqref{9jdecomp} to \eqref{msum} 
to two $9j$-symbols in this expression. A lengthy calculation as above 
results in the last equation in \eqref{Vnpcorrections}.  

\subsection{Mass correction}
$\vev{V_1^2}_c$ would give rise to a $\phi^6$ vertex. 
But we shall see that it is negligible  in the low energy regime. 
In fact, from \eqref{vertices}, $\vev{V_1^2}_c$ becomes 
\begin{align}
\vev{V_1^2}_c=&\sum_{m,m'}\vev{\pon{}\pond{}}_0\trn{N}{\phin^3\ton{}}\trn{N}{\phin^3\tond{}} \nn \\
=&P_N\sum_m(-1)^m\trn{N}{\phin^3\ton{}}\trn{N}{\phin^3\tom{}}.
\end{align}
However, when the momentum $l$ of $\phin$ is much smaller than $2L$, 
that of $\phin^3$ cannot be equal to $2L$ and the traces above vanish by \eqref{Tortho}. 
Hence in the low energy regime of interest in the RG we can neglect this contribution. 

The other remaining terms of ${\cal O}(g_N^2)$ 
will contribute to quadratic terms of $\phin$.
First let us consider $\vev{V_3^2}_c$. Using the trace formula \eqref{4T1}, 
we have after contractions 
\begin{align}
\vev{V_3^2}_c=&P_N^3(2N-1)^3N^4\sum_{(l_1,m_1),(l'_1,m'_1)\in\Li}\pin{1}\pind{1}
\sqrt{(2l_1+1)(2l'_1+1)} \nn \\
&\times\sum_{l,l'}(2l+1)(2l'+1)
\beB
l_1 & 2L & l \\
L & L & L
\eeB
\beB
l'_1 & 2L & l' \\
L & L & L 
\eeB
\beB
l & 2L & 2L \\
L & L & L
\eeB
\beB
l' & 2L & 2L \\
L & L & L 
\eeB \nn \\
&\times
\sum_{m_1,\sim,m_3}\sum_{m,m'}(-1)^{m_1+m_2+m_3-m-m'}
\bep
l_1 & 2L & l \\
m_1 & m_4 & m 
\eep
\bep
l & 2L & 2L \\
-m & m_3 & m_2 
\eep \nn \\
&\times 
\left[
\bep
l'_1 & 2L & l' \\
m'_1 & -m_2 & m'
\eep
\bep
l' & 2L & 2L \\
-m' & -m_3 & -m_4
\eep
+(\text{permutations of}~m_2\sim m_4)
\right].
\end{align}
When $l_1, l'_1\ll L$, the first two $6j$-symbols impose $l=2L-m$, $l'=2L-n$ 
with $m \leq l_1$, $n\leq l_1'$.
Then the Racah formula (\ref{Racah_formula}) and the Stirling's formula 
show that the large-$L$ behavior of the third $6j$-symbol as 
\begin{align}
\beB
2L-m & 2L & 2L \\
L & L & L
\eeB
\simeq \frac{(-1)^{2L-m} 3^{\frac{3}{4}}(2\pi)^\frac14}{8 \sqrt{m!}}L^{\frac{m}{2}-\frac34}\left(\frac34\right)^{3L-\frac{m}{2}},
\label{6jdamp}
\end{align}  
namely, it is exponentially suppressed for $L\gg 1$. Since it is easy to see that other factors in the above equation 
are bounded at least by polynomials of $L$, we conclude that $\vev{V_1^2}_c$ only gives exponentially small 
contribution in the low energy regime. 
Furthermore, it is easy to show that $\vev{V_1V_3}_c=0$ by the momentum conservation. 
Thus only $\vev{V_2V_4}_c$ provides a nonzero contribution even at low
energy. 
From \eqref{vertices} we first find that 
\begin{align}
\vev{V_2^P V_4}_c=P_N^3&\sum_{m_1,\cdots,m_3}(-1)^{m_1+m_2+m_3}\trn{N}{\phin^2\ton{1}\ton{2}} \nn \\
\times\Bigl(&\trn{N}{\tom{2}\ton{1}\ton{}\tom{}}+\trn{N}{\tom{1}\ton{2}\ton{}\tom{}} \nn \\
+&\trn{N}{\tom{2}\ton{}\ton{1}\tom{}}\Bigr),
\end{align}
where the first two terms correspond to planar diagrams, while the
last term to a nonplanar one. 
The formulas \eqref{similarity} and \eqref{coro} enable us to rewrite this as 
\begin{align}
\vev{V_2^P V_4}_c=P_N^3(2N-1)^2N\left(2+(-1)^{2L}
\beB
L & L & 2L \\
L & L & 2L
\eeB
\right)\trn{N}{\phin^2}.
\end{align}
In the second term that comes from the nonplanar diagram,
the $6j$-symbol is again exponentially small as
\begin{align}
\beB
L & L & 2L \\
L & L & 2L
\eeB
\simeq     
\sqrt{\frac{2\pi}{L}} 2^{-4L-2} \,,
\end{align}
for $L\gg 1$.
Therefore for $L\gg 1$ the planar diagram gives 
\begin{align}
\vev{V_2^P V_4}_c\simeq P_N^32(2N-1)^2N\trn{N}{\phin^2}.
\end{align}

Finally, we evaluate $\vev{V_2^{NP} V_4}_c$.
This can be carried out by using the technique explained so far.
We quote only the result,
\begin{align}
    \vev{V_2^{NP} V_4}_c =&
N(2N-1)^2 P_N^3 \trn{N}{
\phin \phinA
-\frac{1}{N} \phin \Lap{ \phinA}
+ \cdots} \,.
\end{align}
Therefore, this term involves the antipode fields.

\section{Antipode transformation and ordering reverse}
\label{app:antipode}

In this appendix we prove a proposition that reveals 
an interesting connection between the ordering of matrices inside the trace 
and the antipode transformation. \\
\underline{Proposition}:
\begin{align}
\trn{N}{\prod_{i=1}^n\phi_i^A}
=\trn{N}{\prod_{i=1}^n\phi_{n+1-i}}.
\label{prop}
\end{align}
\underline{Proof}:\\
The $n=1$ case is trivial. 
The $n=2$ and $n=3$ cases are also obvious because of the
orthogonality (\ref{Tortho}) and the explicit expression of the trace
of three generators (\ref{3T}) with the symmetry property of the $3j$-symbol.
Assuming \eqref{prop} for $n=1,\cdots,k$, 
from \eqref{merging},
\begin{align}
\trn{N}{\prod_{i=1}^{k+1}\phi_i^A}
&=\frac1N\sum_{lm}
\trn{N}{\prod_{i=1}^{k-1}\phi^A_i\tlm}\trn{N}{\phi^A_k\phi^A_{k+1}\tlm^{\dagger}} \nn \\
&=\frac1N\sum_{lm}
\trn{N}{(-1)^l\tlm\prod_{i=1}^{k-1}\phi_{k-i}}\trn{N}{(-1)^l\tlm^{\dagger}\phi_{k+1}\phi_k} \nn \\
&=\frac1N\sum_{lm}
\trn{N}{\prod_{i=1}^{k-1}\phi_{k-i}\tlm}\trn{N}{\phi_{k+1}\phi_k\tlm^\dagger} \nn \\
&=\trn{N}{\phi_{k+1}\phi_k\prod_{i=1}^{k-1}\phi_{k-i}}=\trn{N}{\prod_{i=1}^{k+1}\phi_{k+2-i}},
\end{align}
thus \eqref{prop} holds for $n=k+1$. This completes the induction.



\begin{thebibliography}{99}

\bibitem{Knizhnik:1988ak} 
  V.~G.~Knizhnik, A.~M.~Polyakov and A.~B.~Zamolodchikov,
  ``Fractal Structure of 2D Quantum Gravity,''  
Mod.\ Phys.\ Lett.\ A {\bf 3}, 819 (1988);  

F.~David,
  ``Conformal Field Theories Coupled to 2D Gravity in the Conformal Gauge,''  
Mod.\ Phys.\ Lett.\ A {\bf 3}, 1651 (1988); 
  
J.~Distler and H.~Kawai,
  ``Conformal Field Theory and 2D Quantum Gravity Or Who's Afraid of Joseph Liouville?,''  
Nucl.\ Phys.\ B {\bf 321}, 509 (1989).  


\bibitem{Brezin:1990rb} 
  E.~Brezin and V.~A.~Kazakov,
  ``Exactly Solvable Field Theories Of Closed Strings,''  
Phys.\ Lett.\ B {\bf 236}, 144 (1990);  

M.~R.~Douglas and S.~H.~Shenker,
  ``Strings in Less Than One-Dimension,''  
Nucl.\ Phys.\ B {\bf 335}, 635 (1990);  

D.~J.~Gross and A.~A.~Migdal,
  ``Nonperturbative Two-Dimensional Quantum Gravity,''  
Phys.\ Rev.\ Lett.\  {\bf 64}, 127 (1990);  

M.~R.~Douglas,
  ``Strings In Less Than One-dimension And The Generalized K-d-v Hierarchies,''  
Phys.\ Lett.\ B {\bf 238}, 176 (1990);  


\bibitem{Banks:1996vh} 
  T.~Banks, W.~Fischler, S.~H.~Shenker and L.~Susskind,
  ``M theory as a matrix model: A Conjecture,''  
Phys.\ Rev.\ D {\bf 55}, 5112 (1997)  [hep-th/9610043];  
  
N.~Ishibashi, H.~Kawai, Y.~Kitazawa and A.~Tsuchiya,
  ``A Large N reduced model as superstring,''  
Nucl.\ Phys.\ B {\bf 498}, 467 (1997)  [hep-th/9612115];  
  
  R.~Dijkgraaf, E.~P.~Verlinde and H.~L.~Verlinde,
  ``Matrix string theory,''  
Nucl.\ Phys.\ B {\bf 500}, 43 (1997)  [hep-th/9703030];  
  
J.~M.~Maldacena,
  ``The Large N limit of superconformal field theories and supergravity,''  
Adv.\ Theor.\ Math.\ Phys.\  {\bf 2}, 231 (1998)  [Int.\ J.\ Theor.\ Phys.\  {\bf 38}, 1113 (1999)]  [hep-th/9711200].  


\bibitem{Wilson:1973jj} 
  K.~G.~Wilson and J.~B.~Kogut,
  ``The Renormalization group and the epsilon expansion,''  
Phys.\ Rept.\  {\bf 12}, 75 (1974).  


\bibitem{Brezin:1992yc} 
E.~Brezin and J.~Zinn-Justin,
  ``Renormalization group approach to matrix models,''  
Phys.\ Lett.\ B {\bf 288}, 54 (1992)  [hep-th/9206035].  


\bibitem{Higuchi:1993he} 
S.~Higuchi, C.~Itoi and N.~Sakai,
  ``Exact beta functions in the vector model and renormalization group approach,''  Phys.\ Lett.\ B {\bf 312}, 88 (1993)  [hep-th/9303090];  
  
S.~Higuchi, C.~Itoi, S.~Nishigaki and N.~Sakai,
  ``Nonlinear renormalization group equation for matrix models,''  Phys.\ Lett.\ B {\bf 318}, 63 (1993)  [hep-th/9307116];  

S.~Higuchi, C.~Itoi, S.~Nishigaki and N.~Sakai,
  ``Renormalization group flow in one and two matrix models,''  Nucl.\ Phys.\ B {\bf 434}, 283 (1995)  [Erratum-ibid.\ B {\bf 441}, 405 (1995)]  [hep-th/9409009];  

S.~Higuchi, C.~Itoi, S.~M.~Nishigaki and N.~Sakai,
  ``Renormalization group approach to multiple arc random matrix models,''  Phys.\ Lett.\ B {\bf 398}, 123 (1997)  [hep-th/9612237].  



\bibitem{Hoppe}
J. Hoppe, 
``Quantum Theory of A Massless Relativistic Surface 
and A Two-Dimensional Bound State Problem,'' 
MIT Ph.D.Thesis, 1982. 

B.~de Wit, J.~Hoppe and H.~Nicolai,
 ``On the Quantum Mechanics of Supermembranes,'' 
Nucl.\ Phys.\ B {\bf 305}, 545 (1988).  

J.~Hoppe,
  ``DIFFEOMORPHISM GROUPS, QUANTIZATION AND SU(infinity),'' 
Int.\ J.\ Mod.\ Phys.\ A {\bf 4}, 5235 (1989);  

  J.~Madore,
  ``The Fuzzy sphere,''  
Class.\ Quant.\ Grav.\  {\bf 9}, 69 (1992).  



\bibitem{Martin:2004un}
  X.~Martin,
  ``A Matrix phase for the phi**4 scalar field on the fuzzy sphere,''  JHEP {\bf 0404} (2004) 077  [hep-th/0402230].  

\bibitem{Panero:2006bx}
  M.~Panero,
  ``Numerical simulations of a non-commutative theory: The Scalar
  model on the fuzzy sphere,'' 
JHEP {\bf 0705} (2007) 082  [hep-th/0608202].  

\bibitem{Das:2007gm}
  C.~R.~Das, S.~Digal and T.~R.~Govindarajan,
  ``Finite temperature phase transition of a single scalar field on a
  fuzzy sphere,'' 
 Mod.\ Phys.\ Lett.\ A {\bf 23} (2008) 1781  [arXiv:0706.0695 [hep-th]].  

\bibitem{Steinacker:2005wj}
  H.~Steinacker,
  ``A Non-perturbative approach to non-commutative scalar field theory,''  JHEP {\bf 0503} (2005) 075  [hep-th/0501174].  

\bibitem{Iso:2001mg} 
  S.~Iso, Y.~Kimura, K.~Tanaka and K.~Wakatsuki,
  ``Noncommutative gauge theory on fuzzy sphere from matrix model,''  
Nucl.\ Phys.\ B {\bf 604}, 121 (2001)  [hep-th/0101102].  

\bibitem{Narayan:2002gv} 
  K.~Narayan,
  ``Blocking up D branes: Matrix renormalization?,''  hep-th/0211110.  

\bibitem{Vaidya:2001bt} 
  S.~Vaidya,
  ``Perturbative dynamics on the fuzzy S**2 and RP**2,''  
Phys.\ Lett.\ B {\bf 512}, 403 (2001)  [hep-th/0102212].  

\bibitem{Var}
D.~A.~Varshalovich, A.~N.~Moskalev and V.~K.~Khersonsky,
``Quantum Theory Of Angular Momentum: Irreducible Tensors, 
Spherical Harmonics, Vector Coupling Coefficients, 3nj Symbols,''
{\it  Singapore, Singapore: World Scientific (1988)}.

\bibitem{Kawai:2003yf} 
  H.~Kawai, T.~Kuroki and T.~Morita,
  ``Dijkgraaf-Vafa theory as large N reduction,''  
Nucl.\ Phys.\ B {\bf 664}, 185 (2003)  [hep-th/0303210].


\bibitem{Minwalla:1999px}
  S.~Minwalla, M.~Van Raamsdonk and N.~Seiberg,
  ``Noncommutative perturbative dynamics,''
  JHEP {\bf 0002} (2000) 020
  [hep-th/9912072].

\bibitem{Szabo:2001kg}
  R.~J.~Szabo,
 ``Quantum field theory on noncommutative spaces,''
  Phys.\ Rept.\  {\bf 378} (2003) 207
  [hep-th/0109162]; 

  M.~R.~Douglas and N.~A.~Nekrasov,
 ``Noncommutative field theory,''
  Rev.\ Mod.\ Phys.\  {\bf 73} (2001) 977
  [hep-th/0106048]. 
  
\bibitem{Chu:2001xi} 
  C.~-S.~Chu, J.~Madore and H.~Steinacker,
 ``Scaling limits of the fuzzy sphere at one loop,''  
JHEP {\bf 0108}, 038 (2001)  [hep-th/0106205].  



\bibitem{Azuma:2005bj} 
  T.~Azuma, S.~Bal and J.~Nishimura,
  ``Dynamical generation of gauge groups in the massive Yang-Mills-Chern-Simons matrix model,''  
Phys.\ Rev.\ D {\bf 72}, 066005 (2005)  [hep-th/0504217];  

T.~Aoyama, T.~Kuroki and Y.~Shibusa,
  ``Dynamical generation of non-Abelian gauge group via the improved perturbation theory,''  
Phys.\ Rev.\ D {\bf 74}, 106004 (2006)  [hep-th/0608031].  


\bibitem{Ishii:2008ib} 
  T.~Ishii, G.~Ishiki, S.~Shimasaki and A.~Tsuchiya,
  ``N=4 Super Yang-Mills from the Plane Wave Matrix Model,''  
Phys.\ Rev.\ D {\bf 78}, 106001 (2008) 
 [arXiv:0807.2352 [hep-th]].   

\bibitem{Eguchi:1982nm} 
  T.~Eguchi and H.~Kawai,
  ``Reduction of Dynamical Degrees of Freedom in the Large N Gauge Theory,''  
Phys.\ Rev.\ Lett.\  {\bf 48}, 1063 (1982).  

\bibitem{Kuroki:1998rx} 
  T.~Kuroki,
  ``Master field on fuzzy sphere,''  
Nucl.\ Phys.\ B {\bf 543}, 466 (1999)  [hep-th/9804041].  


\bibitem{VanRaamsdonk:2000rr}
  M.~Van Raamsdonk and N.~Seiberg,
  ``Comments on noncommutative perturbative dynamics,''  JHEP {\bf 0003} (2000) 035  [hep-th/0002186].  


\end{thebibliography}
\end{document}